\documentclass[9pt,twocolumn,twoside,lineno]{pnas-new}
\usepackage{color}
\usepackage{verbatim}
\usepackage{float}
\usepackage{amstext}
\ifx\hypersetup\undefined
  \AtBeginDocument{%
    \hypersetup{unicode=true}
  }
\else
  \hypersetup{unicode=true}
\fi
\usepackage[dot]{bibtopic}

\makeatletter

\title{Multiscale molecular kinetics by coupling Markov state models and
reaction-diffusion dynamics}

\author[a,b,c,d,1]{Mauricio J. del Razo}

\author[d]{Manuel Dibak}

\author[d,e]{Christof Schütte}

\author[d,f,g,1]{Frank Noé}

\affil[a]{Van ’t Hoff Institute for Molecular Sciences, University of Amsterdam,
The Netherlands}

\affil[b]{Korteweg-de Vries Institute for Mathematics, University of Amsterdam,
The Netherlands}

\affil[c]{Dutch Institute for Emergent Phenomena, Amsterdam, The Netherlands}

\affil[d]{Department of Mathematics and Computer Science, Freie Universität
Berlin, Germany}

\affil[e]{Zuse Institute Berlin, Germany}

\affil[f]{Department of Physics, Freie Universität Berlin, Germany}

\affil[g]{Department of Chemistry, Rice University, Houston TX, USA}

\leadauthor{del Razo}

\authorcontributions{M.J.R., C.S. and F.N. designed research; M.J.R. and M.D. performed
research; M.J.R. and F.N. wrote the paper.}

\authordeclaration{The authors declare no conflict of interest.}

\correspondingauthor{\textsuperscript{1} E-mails: \texttt{maojrs@gmail.com and frank.noe@fu-berlin.de}}


\providecommand{\tabularnewline}{\\}
\newcommand{\lyxdot}{.}

\floatstyle{ruled}
\newfloat{algorithm}{tbp}{loa}
\providecommand{\algorithmname}{Algorithm}
\floatname{algorithm}{\protect\algorithmname}

\templatetype{pnasresearcharticle} 

\dates{This manuscript was compiled on \today}
\doi{\url{www.pnas.org/cgi/doi/10.1073/pnas.XXXXXXXXXX}}

\dates{}
\doi{}
\setboolean{displaywatermark}{false}
\let\newmaketitle\maketitle
\let\maketitle\relax

\makeatother

\begin{document}
\maketitle
\begin{abstract}
A novel approach to simulate simple protein-ligand systems at large
time- and length-scales is to couple Markov state models (MSMs) of
molecular kinetics with particle-based reaction-diffusion (RD) simulations,
MSM/RD. Currently, MSM/RD lacks a mathematical framework to derive
coupling schemes; is limited to isotropic ligands in a single conformational
state, and is lacking a multi-particle extensions. In this work, we
address these needs by developing a general MSM/RD framework by coarse-graining
molecular dynamics into hybrid switching diffusion processes. Given
enough data to parametrize the model, it is capable of modeling protein-protein
interactions over large time- and length-scales, and it can be extended
to handle multiple molecules. We derive the MSM/RD framework, and
we implement and verify it for two protein-protein benchmark systems
and one multiparticle implementation to model the formation of pentameric
ring molecules. To enable reproducibility, we have published our code
in the \href{https://github.com/markovmodel/msmrd2}{MSM/RD software package}.
\end{abstract}

\thispagestyle{firststyle}
\ifthenelse{\boolean{shortarticle}}{\ifthenelse{\boolean{singlecolumn}}{\abscontentformatted}{
\twocolumn[\newmaketitle \abscontent] 
}}{}

\section*{I. Introduction}

Molecular dynamics (MD) simulations have allowed the study of a broad
range of biological systems, from small molecules such as anesthetics
or small peptides, to large protein complexes such as the ribosome
or even virus capsids \citep{bernardi2015enhanced}. One of the main
challenges faced by MD simulations is their high computational cost,
which can lead to inadequate sampling of conformational states. While
there is a large body of research focused on sampling of long-time-scale
dynamics of individual macromolecules, there has been less attention
on sampling and simulating the interactions of many macromolecules
on larger length-scales. This is a more complex problem since it not
only involves the long-time dynamics, but it can also involve several
orders of magnitude in length-scale. One landmark example is cellular
signaling, where relevant processes happen across $6$ orders of magnitude
in length-scales (0.1 nm--100 \textmu m) and $18$ orders of magnitude
in time-scales (femtoseconds to hours) \citep{bradshaw2009handbook,hancock2017cell,lim2014cell}.
Two of the most successful approaches to model biomolecular processes
at larger time or length-scales are the following:
\begin{itemize}
\item Markov state models (MSMs) of molecular kinetics are one of the most
well-known techniques to mitigate the MD sampling problem \citep{BowmanPandeNoe_MSMBook,ChoderaNoe_COSB14_MSMs,DoerrEtAl_JCTC16_HTMD,husic2018markov,NoeSchuetteReichWeikl_PNAS09_TPT,PrinzEtAl_JCP10_MSM1}.
They approximate the long-time dynamics of MD systems by Markov chains
on discrete partitions of configuration space. This allows to extract
the long-time kinetics from short MD trajectories and to calculate
molecular observables. State of the art developments have pipelined
the MSM approach into deep learning frameworks \citep{MardtEtAl_VAMPnets,wu2018deep}.
However, larger and more complex systems require sampling an exponentially
growing number of states, constraining its applicability to small
domains with one or a few macromolecules.
\item Particle-based reaction-diffusion (PBRD) simulations are orders of
magnitude more efficient than MD since they model each molecule as
one particle undergoing Brownian diffusion. The solvent effects are
implicitly modeled through the Brownian noise term \citep{hanggi1990reaction,mereghetti2011diffusion,SchoenebergUllrichNoe_BMC14_RDReview},
and the reactions are regulated by reaction rates. For reactions involving
two or more particles, if a pair of reactive particles is close enough
to each other, they can react with a certain reaction rate. They are
ideal to model multi-particle processes at large length-scales but
lack atomic detail. There is a large amount of PBRD literature \citep{agmon1990theory,del2014fluorescence,del2016discrete,del2018grand,dibak2019diffusion,frohner2018reversible,kostre2020coupling,szabo1980first,shoup1982role,schuss2007narrow,szabo1989theory},
as well as several software packages and simulation schemes \citep{andrews2004stochastic,donev2010first,drawert2012urdme,hoffmann2019readdy,takahashi2010spatio,wils2009steps,van2005green,vijaykumar2015combining,vijaykumar2017multiscale,ZonTenWolde_PRL05_GFRD}. 
\end{itemize}
By coupling MSMs of molecular kinetics with PBRD, we can combine the
best of both worlds and perform multiscale molecular simulations across
large time and length-scales; we call this coupling MSM/RD. However,
this coupling is not trivial. The existing implementation of MSM/RD
in \citep{dibak2018msm} suffered from several limitations: there
was no underlying mathematical theory to justify and derive the coupling
scheme; it was limited to simple ligand-protein systems; the protein
was assumed fixed in the frame of reference; the ligand orientation
and possible conformation switching was not taken into account; and
multiparticle extensions were not implemented. However, the aim of
MSM/RD remains the same, to produce efficient multiscale simulations
that reproduce the essential statistical behavior of a practically
unaffordable large-scale MD simulation by employing only statistics
obtained from simulations of the constituent molecules in small solvent
boxes.

In this work, we develop a general framework for MSM/RD that overcomes
the previous shortcomings. It is derived by coarse-graining molecular
kinetics into hybrid switching diffusion processes \citep{mao2006stochastic,yin2010hybrid},
also known as diffusion processes with Markovian switching. These
correspond to a class of stochastic hybrid systems, called ``hybrid''
due to the coexistence of continuous dynamics and discrete events
in the same process. The molecules diffusion corresponds to the continuous
part, while their conformation switching corresponds to the discrete
part. By discretizing the framework, we derive MSM/RD schemes; we
implement and validate them for two protein-protein benchmark systems
and one multiparticle implementation to model the formation of pentameric
ring molecules. Implementations in more realistic systems are left
for future work. However, given enough data to parametrize the model,
the framework is ideal for applications on protein-ligand and protein-protein
dynamics, as well as self-assembly of structures composed of several
copies of the same or a small set of molecules, such as virus capsids.

Previous relevant works \citep{keizer1987diffusion,keizer1982nonequilibrium,keizer1985theory}
introduced spatially dependent reaction rates, a fundamental concept
in our framework and \citep{gopich1999excited,gopich2002kinetics,gopich2016reversible,popov2004influence,szabo1982stochastically}
have modeled fluctuations on the reactivity of the species using Markovian
gates, a special case of our framework. A related multiscale method
\citep{vijaykumar2015combining,vijaykumar2017multiscale} couples
MD with Green's function reaction dynamics, including anisotropic
interactions. Although still constrained by MD computations, this
method could potentially be combined with our approach to accelerate
both MD and PBRD simulations. The work \citep{jagger2020multiscale}
provides an excellent review on several multiscale methods for protein-ligand
binding, including \citep{jagger2018quantitative,votapka2017seekr},
where multiscale simulations are used to estimate kinetic rates. The
ideas presented in this work could help enhance these methods. Moreover,
references \citep{bressloff2015stochastically,bressloff2017stochastic,bressloff2017hybrid,bressloff2018stochastic,bressloff2019protein}
have focused on implementing several stochastic hybrid models in different
fields in biology, which emphasizes the relevance of stochastic hybrid
models in biological settings. Note hybrid switching diffusions are
a general coarse-grained model of MD, and they can thus be applied
to many other applications beyond MSM/RD, such as the diffusion and
conformation switching of molecules under a concentration or temperature
gradient.

\section*{II. Molecular kinetics as hybrid switching diffusions \label{sec:theory_overview}}

\begin{SCfigure*}
\textbf{$\centering$}%
\begin{minipage}[t]{0.6\textwidth}%
\textbf{a.}\includegraphics[width=0.55\textwidth]{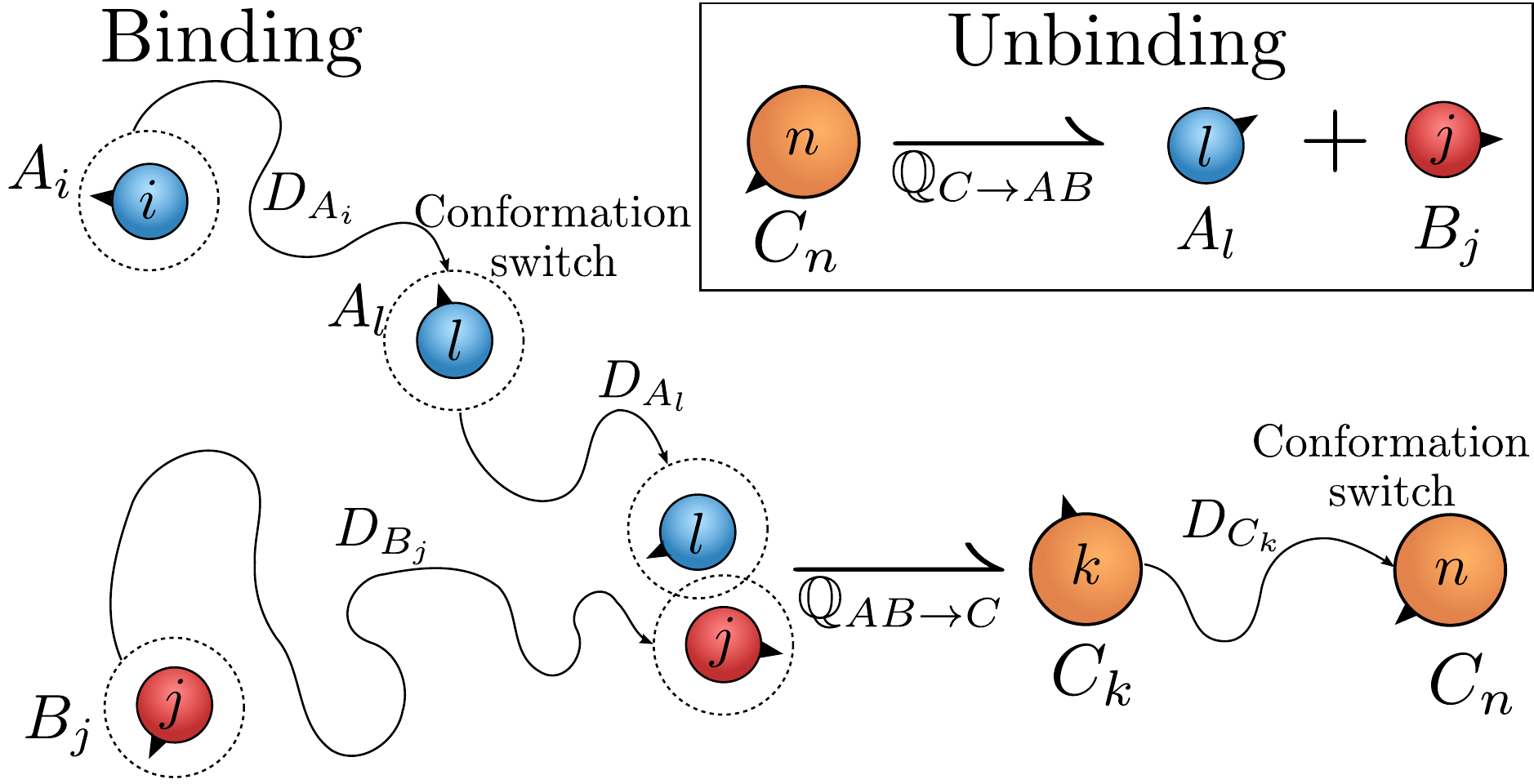}\textbf{
\vline\ c.}\includegraphics[width=0.38\textwidth]{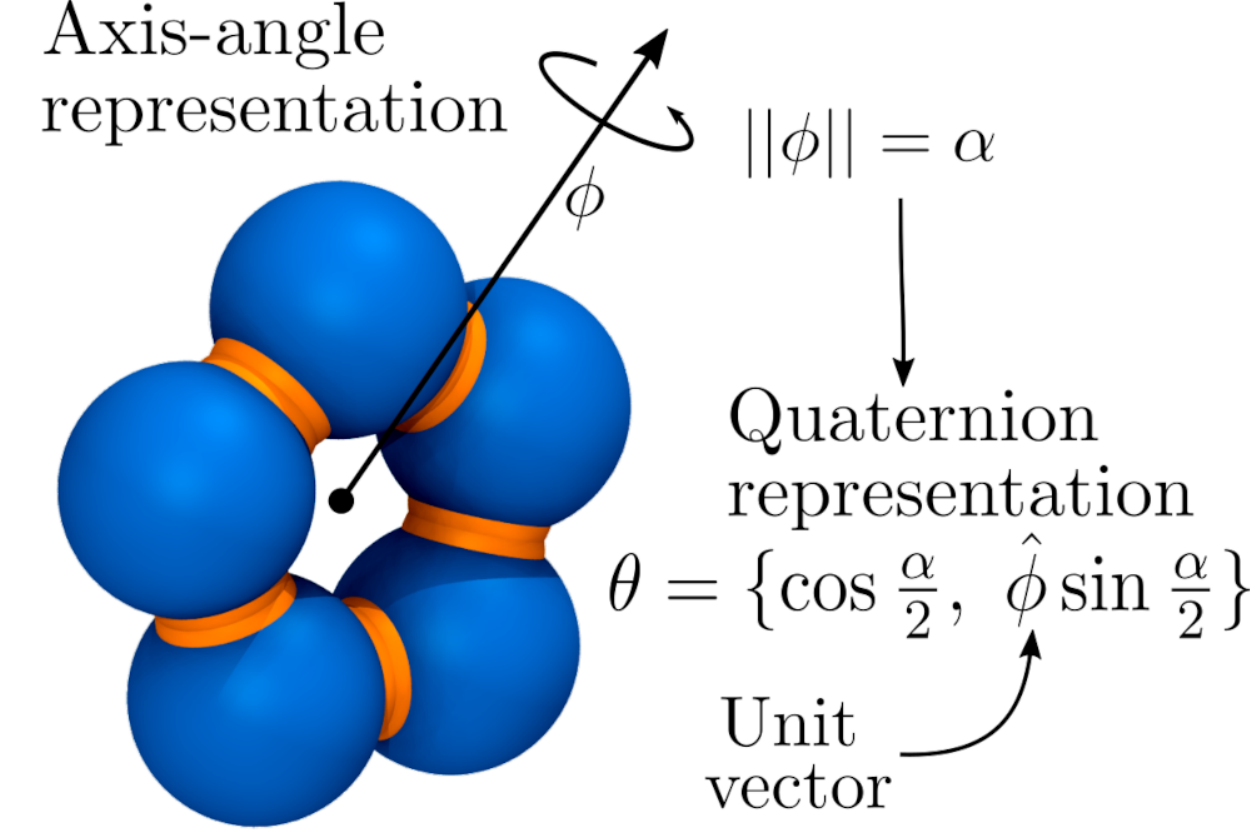}

\medskip{}

\rule[0.5ex]{0.9\columnwidth}{0.5pt}

\medskip{}

\textbf{b.}\includegraphics[width=0.9\textwidth]{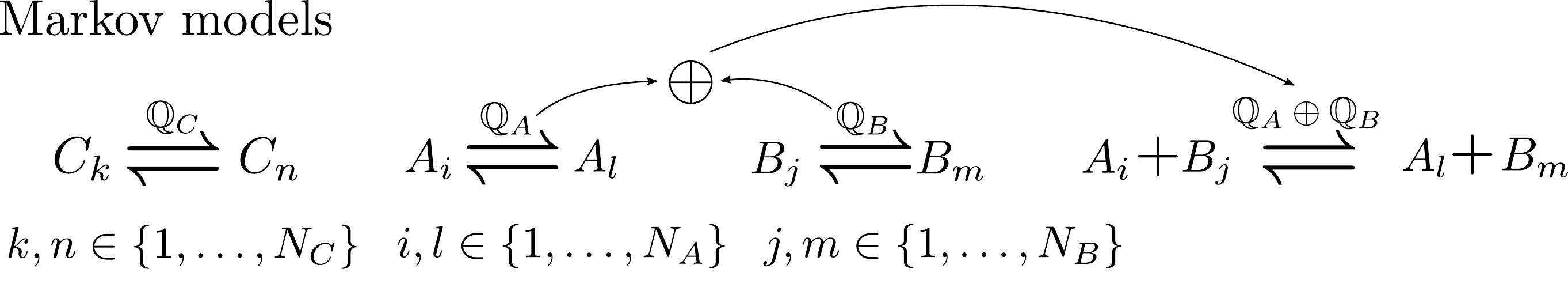}%
\end{minipage}

\caption{Diagrams to illustrate MSM/RD theory and general rotations. \textbf{a.}
Diagram of the binding and unbinding of two reactive molecules, $A+B\rightleftharpoons C$,
when modeling their kinetics as hybrid switching diffusions. Molecules
are represented by particles with position and orientation (black
pointer). The three molecules have a conformation-dependent diffusion,
and the conformations are denoted by a subindex. If molecules $A$
and $B$ are close enough to each other, they transition to a bound
compound $C$ with a configuration-dependent rate given by $\mathbb{Q}_{AB\rightarrow C}$.
The compound $C$ can also unbind into molecules $A$ and $B$ with
a configuration-dependent rate given by $\mathbb{Q}_{C\rightarrow AB}$.
\textbf{b. }Diagram showing the individual Markov models for $C,$$A$
and $B$, and the Markov model for the joint system of molecules $A$
and $B$ when not interacting. \textbf{c.} Orientation of a pentameric
ring molecule using the axis-angle representation with the molecule's
center as reference. The direction of the $\phi$ vector$,$$\hat{\phi}$,
represents the axis of rotation, and its magnitude $||\phi||=\alpha$
represents the radians to be rotated. We can translate this to its
quaternion representation.}

\label{fig:theoryDiags}
\end{SCfigure*}

\subsection*{A. One molecule}

Consider a molecule $A$. If we fix the position and orientation of
the molecule, the position of its atoms only change due to conformational
changes. We can then coarse-grain the all-atom dynamics in configuration
space into an MSM \citep{buchete2008coarse,husic2018markov,PrinzEtAl_JCP10_MSM1}.
Let us assume our molecule $A$ can be described by switching between
two MSM states $A_{1}\rightleftharpoons A_{2}$.

If molecule $A$ is now diffusing instead of being fixed in space,
we would expect different diffusion coefficients in different conformations.
The diffusion and the conformation switching can be modeled together,

\[
\frac{\partial}{\partial t}\left[\begin{array}{c}
p_{1}\\
p_{2}
\end{array}\right]=\left[\begin{array}{c}
D_{1}\nabla^{2}p_{1}\\
D_{2}\nabla^{2}p_{2}
\end{array}\right]+\left(\begin{array}{cc}
-r_{12} & r_{21}\\
r_{12} & -r_{21}
\end{array}\right)\left[\begin{array}{c}
p_{1}\\
p_{2}
\end{array}\right],
\]

where $p(x,t)$ is the vector of probability densities $(p_{1},p_{2})^{T}$
of being in conformation $A_{1}$ or $A_{2}$ at position $x$, $r_{ij}$
are the transition rates from conformation $A_{i}$ to $A_{j}$ that
form the corresponding transition rate matrix. Note the first term
of the right hand side corresponds to the Fokker-Planck equations
of the diffusion processes, while the second term corresponds to a
continuous-time MSM, or Master-equation model \citep{buchete2008coarse}.
We would like to incorporate rotational diffusion and generalize it
to $N$ different conformations. The resulting generalization yields

\begin{equation}
\frac{\partial p(x,t)}{\partial t}=\underbrace{\mathcal{D}p(x,t)}_{\text{Diffusion}}+\underbrace{\mathbb{Q}p(x,t)}_{\text{MSM}},\label{eq:diffMS_01}
\end{equation}
where $p(x,t)=(p_{1},\dots,p_{N})^{T}$ is the vector of probability
densities of being in the corresponding conformations at $x$ and
time $t$, with $x=(r,\theta)$ denoting the position and orientation
coordinates of the molecule. The operator $\mathcal{D}$ describes
the translational and rotational diffusion of the molecule in each
of its conformations. The matrix $\mathbb{Q}$ is a $N\times N$ transition
rate matrix describing the conformation switching; its diagonal entries
are all negative and its non-diagonal ones positive; its columns sum
to zero. In this way, the equation models simultaneously the molecule's
diffusion and the switching of conformation. Equation \ref{eq:diffMS_01}
is an example of a hybrid switching diffusion process, and one could
also write a stochastic differential equation for the individual stochastic
trajectories. A detailed derivation of this theory is presented in
the SI Appendix \ref{app:fullTheory}. The diffusion operator and
the transition rate matrix can be a function of $x$, $\mathcal{D}(x)$
and $\mathbb{Q}(x)$, which provides a robust framework for several
interesting applications. In this work, we are interested in the interaction
between two molecules, so we generalize this result for two interacting
molecules.

\subsection*{B. Two interacting molecules}

Consider two molecules $A$ and $B$. If they are far enough from
each other, they will not interact. Each molecule has a state vector
assigned, $p_{A}$ and $p_{B}$, with sizes $N_{A}$ and $N_{B}$
corresponding to their respective number of conformations. The conformations
are denoted by $A_{i}$ and $B_{j}$ with $i={1,\dots,N_{A}}$ and
$j={1,\dots,N_{B}}.$ The diffusion operators $\mathcal{D}_{A}$ and
$\mathcal{D}_{B}$ encode the rotational and translational diffusion,
which in the simplest case will correspond to Laplacian operators
with diffusion coefficients for the different conformations, $D_{A_{i}}$
and $D_{B_{j}}.$ The rate matrices $\mathbb{Q}_{A}$ and $\mathbb{Q}_{B}$
encode the rates at which they switch conformation. Each molecule
will satisfy its own version of Eq. \ref{eq:diffMS_01}

\begin{equation}
\frac{\partial p_{A}}{\partial t}=\mathcal{D}_{A}p_{A}+\mathbb{Q}_{A}p_{A},\qquad\frac{\partial p_{B}}{\partial t}=\mathcal{D}_{B}p_{B}+\mathbb{Q}_{B}p_{B},\label{eq:diffMS_two}
\end{equation}
see Fig. \ref{fig:theoryDiags}a for a graphical reference. The state
of the system $p_{AB}$ is given by all the possible combinations
of states of $A$ and states of $B.$ This corresponds to the tensor
product of all the states of $A$ with all the states of $B$, i.e.
$p_{AB}=p_{A}\otimes p_{B}$ \citep{hempel2021independent}. For instance,
if $A$ and $B$ have two states each, $A_{1},\,A_{2}$ and $B_{1},\,B_{2}$
respectively, the full system given by the tensor product has four
possible states: $A_{1}B_{1}$, $A_{1}B_{2}$, $A_{2}B_{1}$ and $A_{2}B_{2}$.
Taking the time derivative of $p_{A}\otimes p_{B}$ and using Eqs.
\ref{eq:diffMS_two}, we obtain

\begin{equation}
\frac{\partial p_{AB}(x)}{\partial t}=\mathcal{D}p_{AB}(x)+\left(\mathbb{Q}_{A}\oplus\mathbb{Q}_{B}\right)p_{AB}(x),\label{eq:diffMS_02_ind}
\end{equation}
where the diffusion operator is applied independently before taking
the tensor product $\mathcal{D}p_{AB}=\mathcal{D}_{A}p_{A}\otimes p_{B}+p_{A}\otimes\mathcal{D}_{B}p_{B}$
and $\mathbb{Q}_{A}\oplus\mathbb{Q}_{B}=\left(\mathbb{Q}_{A}\otimes\mathbb{I}_{N_{B}}\right)+\left(\mathbb{I}_{N_{A}}\otimes\mathbb{Q}_{B}\right)$
is the Kronecker sum with $\mathbb{I}_{K}$ the identity matrix of
order $K$. The appearance of the Kronecker sum results evident when
computing the solution of the full system as the tensor product of
the individual solutions of Eqs. \ref{eq:diffMS_two}, $p_{AB}(x,t)=e^{t\mathcal{D}_{A}}\otimes e^{t\mathcal{D}_{B}}\otimes e^{t(\mathbb{Q}_{A}\oplus\mathbb{Q}_{B})}p_{AB}(x,0)$
\citep{hempel2021independent}. This means that the rate matrix of
the full system is given by the transition rate matrix $\mathbb{Q}_{A}\oplus\mathbb{Q}_{B}$
(Fig. \ref{fig:theoryDiags}b). Note that if we were using a discrete-time
MSM, the transition probability matrix of the full system will simply
be the tensor product of the independent transition probability matrices.

Let's assume now molecules $A$ and $B$ are close to each other and
are interacting, such that they can be considered as a complex $C$
that diffuses as a single entity. The state vector is $p_{C}$ with
dimension $N_{C}$, corresponding to the bound conformations, $C_{k}$
with $k={1,\dots,N_{C}}$. We can thus write it in the form of Eq.
\ref{eq:diffMS_two}

\begin{equation}
\frac{\partial p_{C}}{\partial t}=\mathcal{D}_{C}p_{C}+\mathbb{Q}_{c}p_{C}.\label{eq:diffMS_02_cou}
\end{equation}

We would like to switch smoothly between the non-interacting regime
(Eq. \ref{eq:diffMS_02_ind}) and the bound regime (Eq. \ref{eq:diffMS_02_cou}),
so we introduce a transition regime, where the molecules are still
dissociated but interacting, and the transition rates strongly depend
on the relative position and orientation between the molecules. The
dynamics of the system in the three regimes can be written in terms
of the probability of being in any of the dissociated states (AB)
(Eqs. \ref{eq:diffMS_02_ind}) and any of the bound states (C) (\ref{eq:diffMS_02_cou}),
namely $p(x,t)=(p_{AB},p_{C})^{T}$, and a transition rate matrix
$\mathbb{Q}(x)$ that depends on the phase space coordinates $x$,
more specifically on the relative position and orientation between
the molecules. We can write the dynamics of $p(x,t)$ as a hybrid
switching diffusion process

\begin{equation}
\frac{\partial p(x)}{\partial t}=\mathcal{D}p(x)+\mathbb{Q}(x)p(x),\:\begin{array}{l}
\mathbb{Q}(x)=\left(\begin{array}{c|c}
\mathbb{Q}_{AB} & \mathbb{Q}_{C\rightarrow AB}\\
\hline \mathbb{Q}_{AB\rightarrow C} & \mathbb{Q}_{C}
\end{array}\right).\end{array}\label{eq:diffMS_MSMRD}
\end{equation}
The matrix $\mathbb{Q}_{AB\rightarrow C}$ contains the transition
rates from dissociated states (AB) to bound states ($C$), and vice
versa for the matrix $\mathbb{Q}_{C\rightarrow AB}$. If the initial
relative distance between the molecules is large enough, the molecules
are in the non-interacting regime; $\mathbb{Q}_{AB\rightarrow C}$
is zero; the system can only reach the states accessible by $\mathbb{Q}_{A}\oplus\mathbb{Q}_{B}$
--so $\mathbb{Q}_{AB}=\mathbb{Q}_{A}\oplus\mathbb{Q}_{B}$-- and
the dynamics given by Eq. \ref{eq:diffMS_02_ind} are recovered. However,
diffusion can bring the molecules together into the transition regime,
making $\mathbb{Q}_{AB\rightarrow C}$ nonzero and allowing the system
to transition to the bound regime. The system can then transition
to other bound states through $\mathbb{Q}_{C}$, and it can transition
out of the bound regime into the transition regime (dissociated) through
$\mathbb{Q}_{C\rightarrow AB}$. Note columns of $\mathbb{Q}(x)$
should sum to zero for any given $x$, and $\mathbb{Q}_{C}$ on Eq.
\ref{eq:diffMS_MSMRD} is a renormalized version of $\mathbb{Q}_{c}$
in Eq. \ref{eq:diffMS_02_cou}.

Equation \ref{eq:diffMS_MSMRD} constitutes the general MSM/RD framework,
and its dynamics are represented in Figs. \ref{fig:theoryDiags}a
and \ref{fig:theoryDiags}b. The SI Appendix \ref{app:twomolCD} shows
a more detailed derivation of this theory. Discretizations of this
model are used to generate MSM/RD schemes. In the methods section,
we derive the MSM/RD schemes used throughout this work by doing a
piecewise constant discretization of $\mathbb{Q}(x)$. Their parametrization
and explicit algorithms are given in the SI appendices \ref{app:paramMSM/RD}
and \ref{app:MSM/RD-schemes}.

\subsection*{C. Quaternions \label{app:Quaternions}}

The MSM/RD framework requires a representation for the orientation
or rotation $\theta$ of a rigid body, such as: Euler angles, rotation
matrices or unit quaternions among others. Some of these have severe
disadvantages, such as the gimbal lock in Euler angles, while unit
quaternions have proved to be the most simple, robust and numerically
efficient \citep{delong2015brownian,linke2018fully,rapaport1985molecular,vijaykumar2017multiscale}.
A quaternion $\theta=\{s,p\}$, consists of a real part $s$ and a
three-dimensional vector part $p.$ If normalized to one, $s^{2}+p\cdot p=1$,
it can be used to represent a three-dimensional rotation. Let us consider
first a more physically intuitive representation of rotations, the
axis-angle representation, where an arbitrary rotation is represented
by a three dimensional vector $\phi$ (Fig. \ref{fig:theoryDiags}c).
Its direction $\hat{\phi}=\phi/\left\Vert \phi\right\Vert $ corresponds
to the axis of rotation following the right hand rule, and the length
of the vector $\left\Vert \phi\right\Vert $ corresponds to the magnitude
of rotation (Fig. \ref{fig:theoryDiags}c). The corresponding quaternion
associated to this rotation is

\begin{equation}
\theta=\{\cos\left(\left\Vert \phi\right\Vert /2\right),\;\sin\left(\left\Vert \phi\right\Vert /2\right)\hat{\phi}\}.\label{eq:axangle2quaternion}
\end{equation}

Similar to complex numbers, quaternions are further endowed with an
algebraic structure such that the resulting rotation of consecutive
rotations, $\theta_{1}$ and $\theta_{2}$, is obtained by an algebraic
multiplication, $\theta=\theta_{2}\theta_{1}$,

\[
\theta=\{s_{2}s_{1}-p_{2}p_{1},\;s_{2}p_{1}+s_{2}p_{1}+p_{2}\times p_{1}\},
\]

where the cross product makes the multiplication non-commutative,
as expected for rotations. The unit quaternion $\theta^{-1}=\{s,-p\}$
is the inverse quaternion of $\theta$ representing the inverse rotation,
such that $\theta\theta^{-1}=I$ is the identity rotation. Note the
quaternion $-\theta$ corresponds to the same rotation as $\theta$;
therefore it is enough to use half of the surface of the four-dimensional
unit sphere to describe all possible rotations in three-dimensional
space. If a one to one relation is desired, the simplest choice is
to restrict to $s\geq0$. More detailed accounts of quaternions can
be found in the literature \citep{delong2015brownian,rapaport1985molecular,vijaykumar2017multiscale}.

\bigskip{}

\section*{III. Methods}

\subsection*{A. A general MSM/RD scheme \label{sec:MSMRDscheme}}

\begin{figure*}
\centering \textbf{a.}\includegraphics[width=0.38\textwidth]{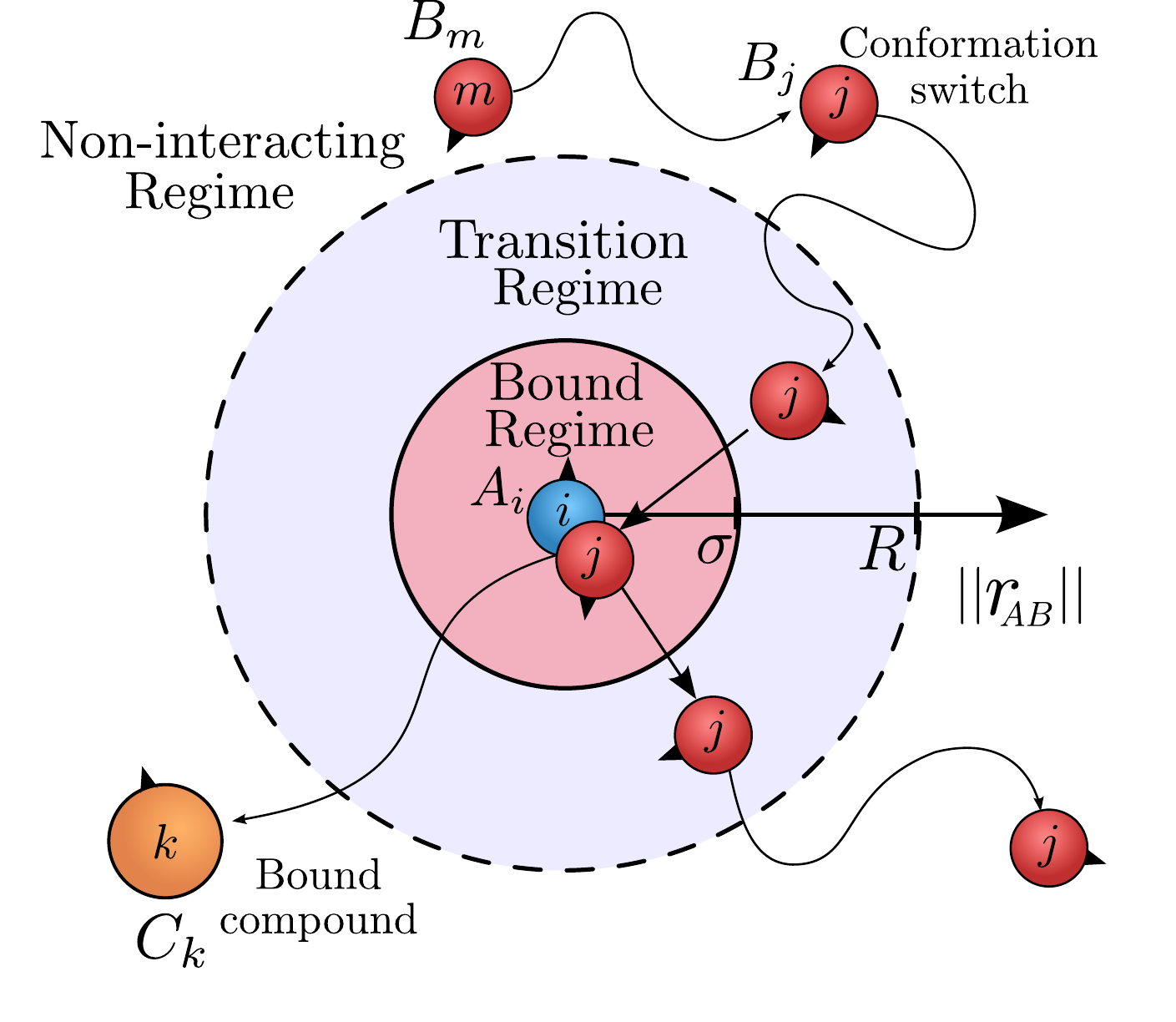}\textbf{b.}\includegraphics[width=0.58\textwidth]{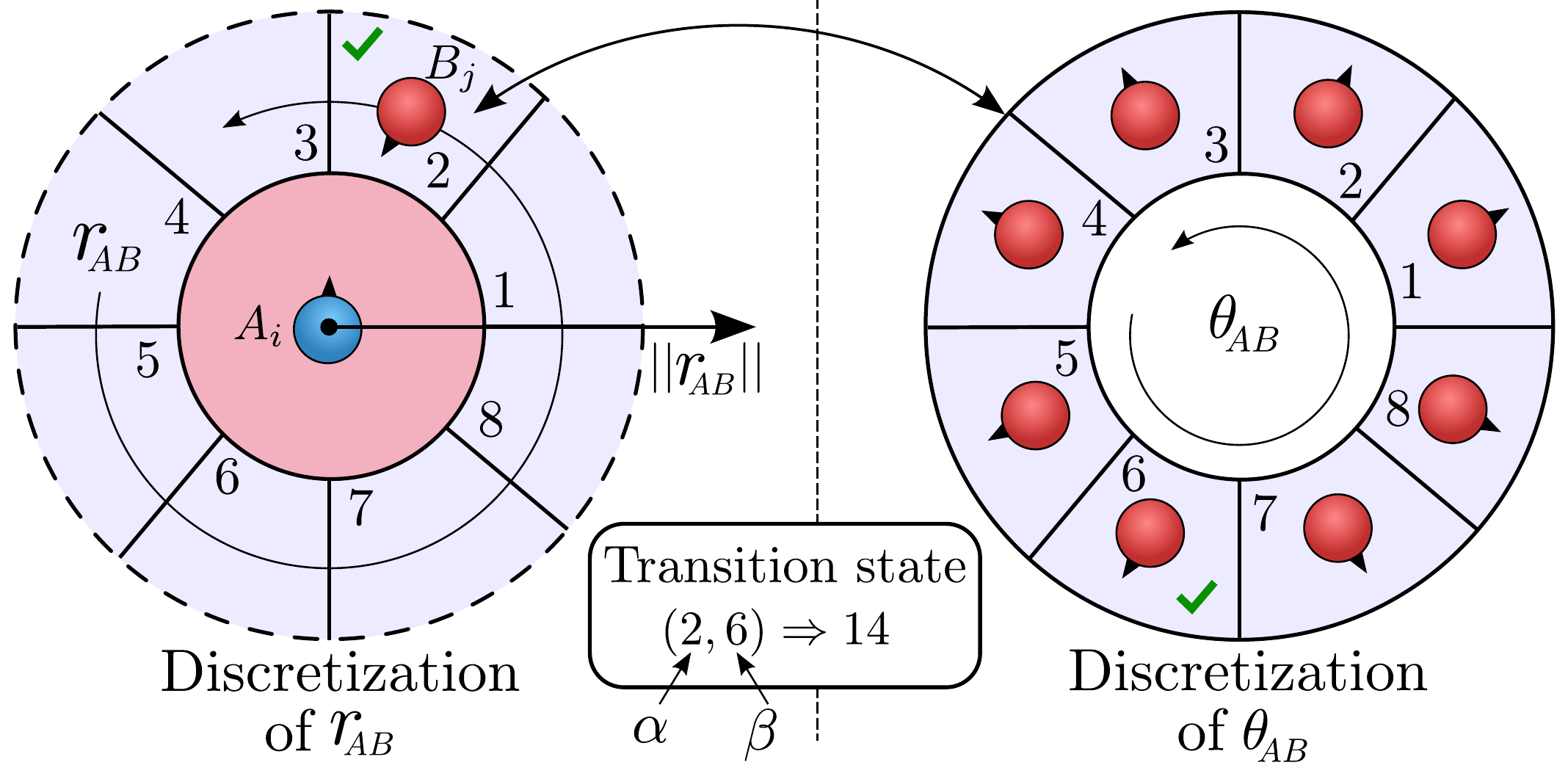}\caption{Discretization diagrams for the MSM/RD scheme\textbf{ a. }Diagram
of the three different regimes in the MSM/RD scheme defined by $\sigma<\left\Vert r_{AB}\right\Vert <R$.
The bound regime is shaded in red; the transition regime is shaded
in blue and the non-interacting regime is in white. To define these
regions, we fix the frame of reference to molecule $A$. In the non-interacting
regime, they both diffuse and change conformation freely. In the transition
regime, they can transition to a bound compound state. From the bound
state, they can unbind and switch to a specific configuration in the
transition regime. From the transition regime, they can diffuse away
into the non-interacting regime. Note the orientation is specified
by a small black pointer attached to each particle.\textbf{ b.} Definition
of transition states (or unbound transition states) within the transition
regime. To define the transition states, we perform two discretizations:
one for the relative position $r_{AB}$ and one for the relative orientation
$\theta_{AB}$. In this illustration, the relative position is simply
a two dimensional vector, so we simply partition the blue shaded ring
($\sigma<\left\Vert r_{AB}\right\Vert <R$) from $0$ to $2\pi$ into
$8$ parts. For each discrete value of the relative position, the
molecules can still have an arbitrary relative orientation, so, we
also need to discretize the relative orientation. The relative orientation
is represented with one degree of freedom, so we discretize it by
partitioning the circle into $8$ parts. If $\alpha$ is the discretization
state of $r_{AB}$ and $\beta$ the one of $\theta_{AB},$ the transition
state number is given by $(\alpha-1)8+\beta$. This discretization
yields a total of $8\times8=64$ transition states. In each transition
state, the rates are approximated by a constant value, yielding a
piecewise constant approximation of the rates in $\mathbb{Q}$. In
our MSM/RD implementation, an analogous discretization is done in
three dimensions (six degrees of freedom).}

\label{fig:discretization}
\end{figure*}

The general MSM/RD framework for two interacting molecules is condensed
in Eq. \ref{eq:diffMS_MSMRD}. In most cases, we will not know the
rate functions constituting Eq. \ref{eq:diffMS_MSMRD}. However, we
can discretize the equation and obtain a specific MSM/RD coupling
scheme, which can be parametrized with MD trajectories. Equation \ref{eq:diffMS_MSMRD}
thus provides a robust theoretical foundation from which different
MSM/RD schemes can be derived by applying different discretizations;
it serves as a guideline to derive different and better suited schemes
for the situation at hand.

The MSM/RD schemes used throughout this work originate from piecewise
constant discretizations of the transition rate matrix $\mathbb{Q}(x)$
from Eq. \ref{eq:diffMS_MSMRD}. We first divide the phase space in
the three main regions/regimes: non-interacting, transition and bound
regimes (Fig. \ref{fig:discretization}a). The definition of these
regimes will be system dependent and based on the relative position
between the two molecules. As a rule of thumb, the interaction between
molecules must be weak in the transition regime and effectively zero
in the non-interacting regime. MSM/RD requires parametrizing two MSMs,
one for the non-interacting regime, $\mathbb{Q}_{A}\oplus\mathbb{Q}_{B}$,
and one for the transition and bound regime together. In the sections
below, we will cover how the MSM/RD dynamics are constructed in each
of these regions. In the SI Appendices \ref{app:MSM/RD-schemes} and
\ref{app:paramMSM/RD}, we further show the corresponding MSM/RD algorithm
and how to discretize the MD trajectories to parametrize the MSM/RD
scheme.

\subsubsection*{Non-interacting regime}

We consider two molecules $A$ and $B$ as rigid bodies with relative
position $r_{AB}=r_{B}-r_{A}$, and relative orientation $\theta_{AB}=\theta_{B}\theta_{A}^{-1}$,
where $\theta_{A}$ and $\theta_{B}$ are quaternions representing
orientations. Following Eq. \ref{eq:diffMS_two}, if the two molecules
are far enough apart, $\left\Vert r_{AB}\right\Vert \geq R$, they
diffuse and change conformation independently. Thus the rates of the
transition matrix $\mathbb{Q}(x)$ do not depend on $r_{AB}$ or $\theta_{AB}$,
and the dynamics of the individual molecules are discretized into
individual MSMs using standard methods \citep{PrinzEtAl_JCP10_MSM1},
yielding $\mathbb{Q}_{AB}=\mathbb{Q}_{A}\oplus\mathbb{Q}_{B}$. For
the sake of simplicity and without loss of generality, we assume the
particles are modeled with overdamped Langevin dynamics. The corresponding
SDE based on \ref{eq:diffMS_SDE} is

\begin{equation}
\frac{dX_{k}(\eta_{k},t)}{dt}=\sqrt{2k_{B}T}M_{k}^{\frac{1}{2}}(\eta_{k})\xi(t),\label{eq:overdGeneral}
\end{equation}

where $k$ denotes the molecule $A$ or $B$; $dX_{k}=[dr_{k},\,d\Phi_{k}]$,
with $dr_{k}$ the change in position of molecule $k$ and $d\Phi_{k}$
its change of orientation in the axis-angle representation; $M_{k}$
the mobility matrix of molecule $k$, which depends on its conformation
$\eta_{k}$; and $\xi(t)$ corresponds to six-dimensional Gaussian
white noise. The conformation $\eta_{k}$ of each molecule changes
following a discrete- or continuous-time MSM with the constant rates
from the transition matrix $\mathbb{Q}_{A}\oplus\mathbb{Q}_{B}$.
Thus, $\eta_{k}$ can be propagated by simply sampling transition
probabilities in the discrete case or by using a Gillespie-type algorithm\citep{anderson2015stochastic,gillespie2007stochastic}
in the continuous case. This description corresponds to the trajectory
representation of the stochastic process described by Eq. \ref{eq:diffMS_two}.
If the translational and rotational motion are weakly coupled and
both isotropic, we can approximate Eq. \ref{eq:overdGeneral} by
\begin{align}
\frac{dr_{k}(\eta_{k},t)}{dt} & =\sqrt{2D_{k}(\eta_{k})}\xi(t),\label{eq:ovedSplit}\\
\frac{d\Phi_{k}(\eta_{k},t)}{dt} & =\sqrt{2D_{k}^{\text{rot}}(\eta_{k})}\xi_{\text{rot}}(t),\nonumber 
\end{align}

where $D_{k}$ and $D_{k}^{\text{rot}}$are the translation and rotational
diffusion coefficients of molecule $k$, and in these equations, $\xi(t)$
and $\xi_{\text{rot}}(t)$ each correspond to three-dimensional Gaussian
white noise. Note in this region the $C$ states are not accessible,
so only $\mathbb{Q}_{AB}$ is relevant. The numerical discretization
of this equation has the same form as Eq. \ref{eq:euler-maruyama}
but with zero force and torque terms. The diffusion coefficients (or
matrices in the general case) should also be estimated from MD trajectories.
There are several works focused on this topic \citep{bullerjahn2020optimal,linke2018fully,qian1991single};
we also added a small section about it in the SI Appendix \ref{app:diffusionEstimation}.

\subsubsection*{Transition regime}

The transition regime is defined by the region between the non-interacting
and the bound regime, $\sigma<\left\Vert r_{AB}\right\Vert <R$. In
this regime, the transition rates depend continuously on the relative
position and orientation of the molecules. As we plan to infer these
rates from MD simulations, it is convenient to discretize $\mathbb{Q}(x)$
into a relatively small number of transition regions/states where
the rates are approximated by constant values, yielding a piecewise
constant approximation of $\mathbb{Q}(x)$ that is easier to infer
from MD data.

Figure \ref{fig:discretization} shows an illustration of the different
regions/states and the discretization of the transition regime for
a simplified lower dimensional case. For each transition state, given
by the combination of a discrete value of the relative position and
of the relative orientation, we approximate the rates in $\mathbb{Q}(x)$
by a constant value, yielding a piecewise constant approximation of
$\mathbb{Q}_{AB}$ and $\mathbb{Q}_{AB\rightarrow C}$ in the transition
regime. In this regime the particles are always dissociated, so $\mathbb{Q}_{C\rightarrow AB}$
and $\mathbb{Q}_{C}$ are not relevant.

In general, the relative position and the relative orientation account
for a total of six degrees of freedom, so the discretization of the
transition regime is much more complex than in Fig. \ref{fig:discretization},
but it still follows the same principle. The first step is to provide
an equal area partition of the surface of the sphere following \citep{leopardi2006partition},
yielding a discretization of the relative position in the transition
region. Then, we need to discretize the relative orientation, which
is given in terms of a unit quaternion. As unit quaternions can be
projected into the top half three dimensional unit sphere, we use
the same equal area partition sphere with a few additional cuts along
the radial direction, yielding an effective discretization of all
the possible relative orientations. It is important to keep the number
of divisions in these partitions as small as possible to avoid an
exploding number of transition states.

Note conformation switching within the transition regime is naturally
incorporated in the framework. The transition matrix $\mathbb{Q}(x)$
acts on $p(x,t)=(p_{AB},p_{C})^{T}$, where $p_{AB}$ includes one
entry for every possible conformation combination between the two
molecules. Thus, the discretization of $\mathbb{Q}(x)$ includes the
rates corresponding to conformation changes within the transition
regime. Alternatively, by collapsing all conformations into one state
in the parametrization, one can obtain averaged rates over all conformations
for all the transitions within the transition regime.

In the transition regime, the diffusion of the particles ---approximated
by Eq. \ref{eq:overdGeneral}--- and the propagation of the MSM ---following
$\mathbb{Q}_{AB}$ and $\mathbb{Q}_{AB\rightarrow C}$--- are run
in parallel. If $\left\Vert r_{AB}\right\Vert $ becomes larger than
$R$ due to diffusion, the MSM is ignored and the dynamics switch
to the non-interacting regime. If a binding event occurs, the diffusion
of the binding particle is ignored and the dynamics switch to the
bound regime.

\subsubsection*{Bound regime}

If molecules $A$ and $B$ are close enough to each other, $\left\Vert r_{AB}\right\Vert \leq\sigma$,
they are strongly interacting and can be considered as a bound compound
$C$ with several metastable configurations. In this case, their diffusion
and conformation switching are no longer independent, and the transition
rates do not depend on $r_{AB}$ and $\theta_{AB}$, so they are assumed
constant. The transitions in the bound regime can be between metastable
states (following $\mathbb{Q}_{C}$) or towards an unbound state in
the transition regime (following $\mathbb{Q}_{C\rightarrow AB}$).
Analogously to the previous example, we assume without loss of generality
that the dynamics of the compound follow overdamped Langevin dynamics
\begin{equation}
\frac{dX_{C}(\eta_{C},t)}{dt}=\sqrt{2k_{B}T}M_{C}^{\frac{1}{2}}(\eta_{C})\xi(t),\label{eq:overdBound}
\end{equation}

with $dX_{C}=[dr{}_{C},\,d\Phi_{C}]$. The conformation $\eta_{C}$
is propagated using the rates from $\mathbb{Q}_{C}$ and $\mathbb{Q}_{C\rightarrow AB}$.
The dynamics are propagated in the same way as Eq. \ref{eq:overdGeneral}.
If the translational and rotational motion are weakly coupled and
isotropic, we can obtain analogous results to that of Eq. \ref{eq:ovedSplit}
with $k=C$ and with analogous numerical discretization and diffusion
coefficient estimation. If a transition towards a dissociated state
in the transition regime happens, the dynamics switch to the transition
regime. In the bound regime, particles are always bound, so $\mathbb{Q}_{AB\rightarrow C}$
and $\mathbb{Q}_{AB}$ are not relevant. Note when parametrizing $\mathbb{Q}(x)$
from MD data, we obtain one MSM at once for both the transition and
bound regime (SI appendix \ref{app:paramMSM/RD}), which describes
all states in which $A$ and $B$ are interacting, including strongly
and weakly bound states, intermediates between unbound and bound state
and even dissociated states in which $A$ and $B$ are sufficiently
close to induce a force upon each other. Such MSMs have, for example,
been computed for protein-ligand and protein-protein association in
the past few years \citep{BuchFabritiis_PNAS11_Binding,SilvaHuang_PlosCB_LaoBinding,PlattnerNoe_NatComm15_TrypsinPlasticity,PlattnerEtAl_NatChem17_BarBar}.

\subsection*{B. Benchmark MD model: patchy particles \label{subsec:benchmarkMD}}

\begin{figure*}
\centering \textbf{a.}\includegraphics[width=0.41\textwidth]{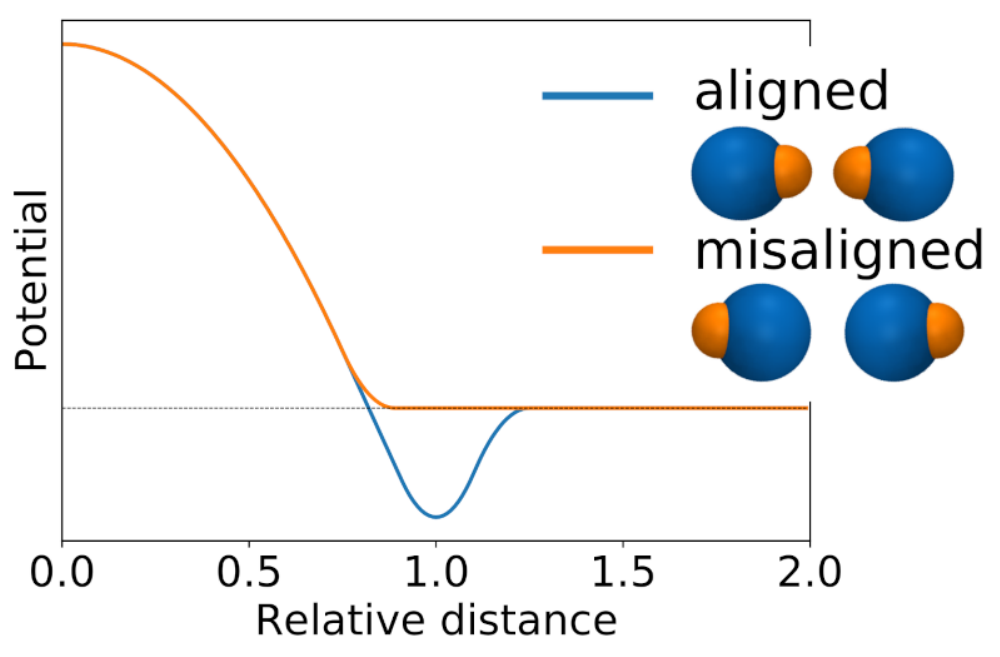}\textbf{b.\includegraphics[width=0.39\textwidth]{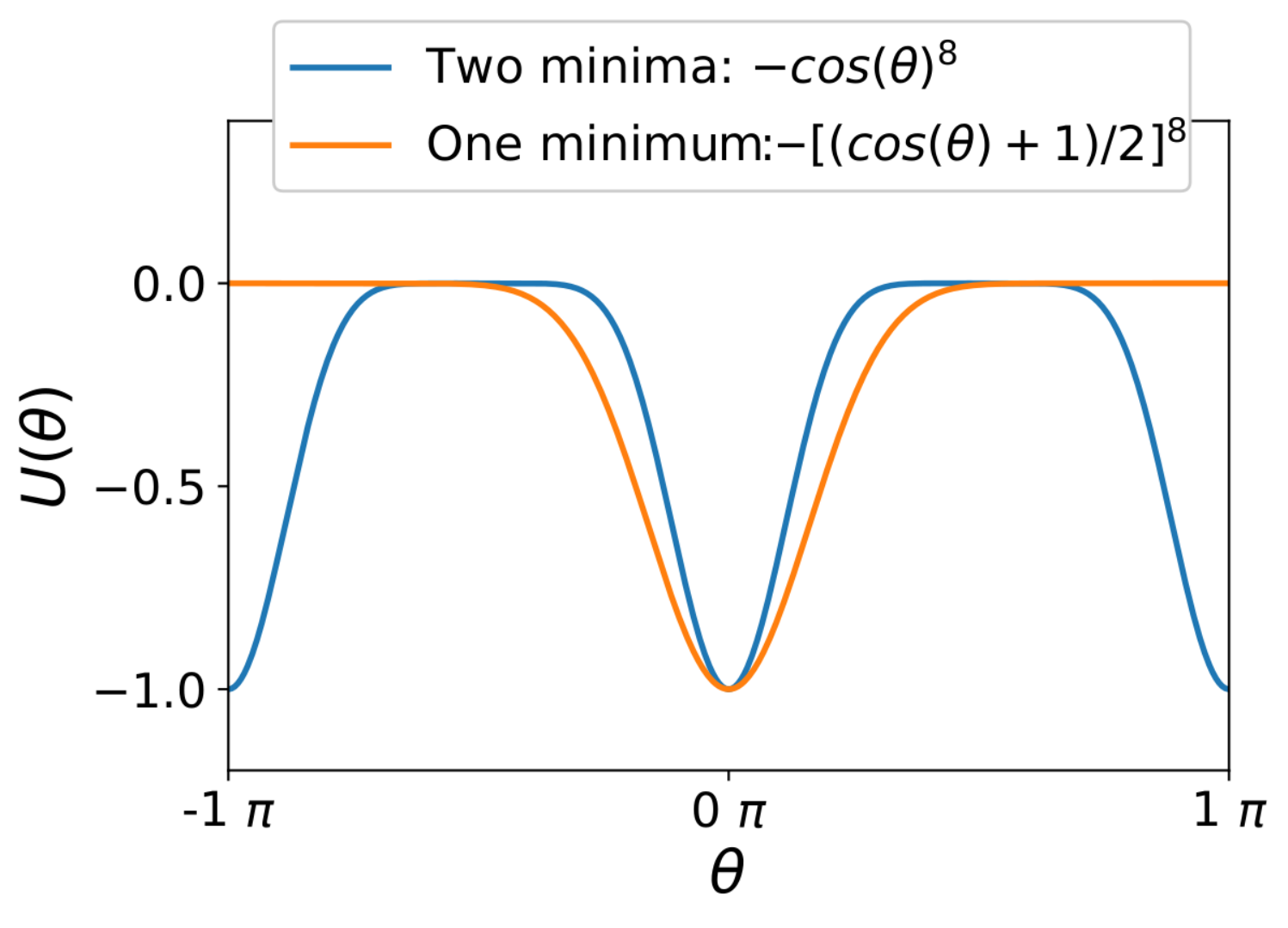}c.}\includegraphics[width=0.12\textwidth]{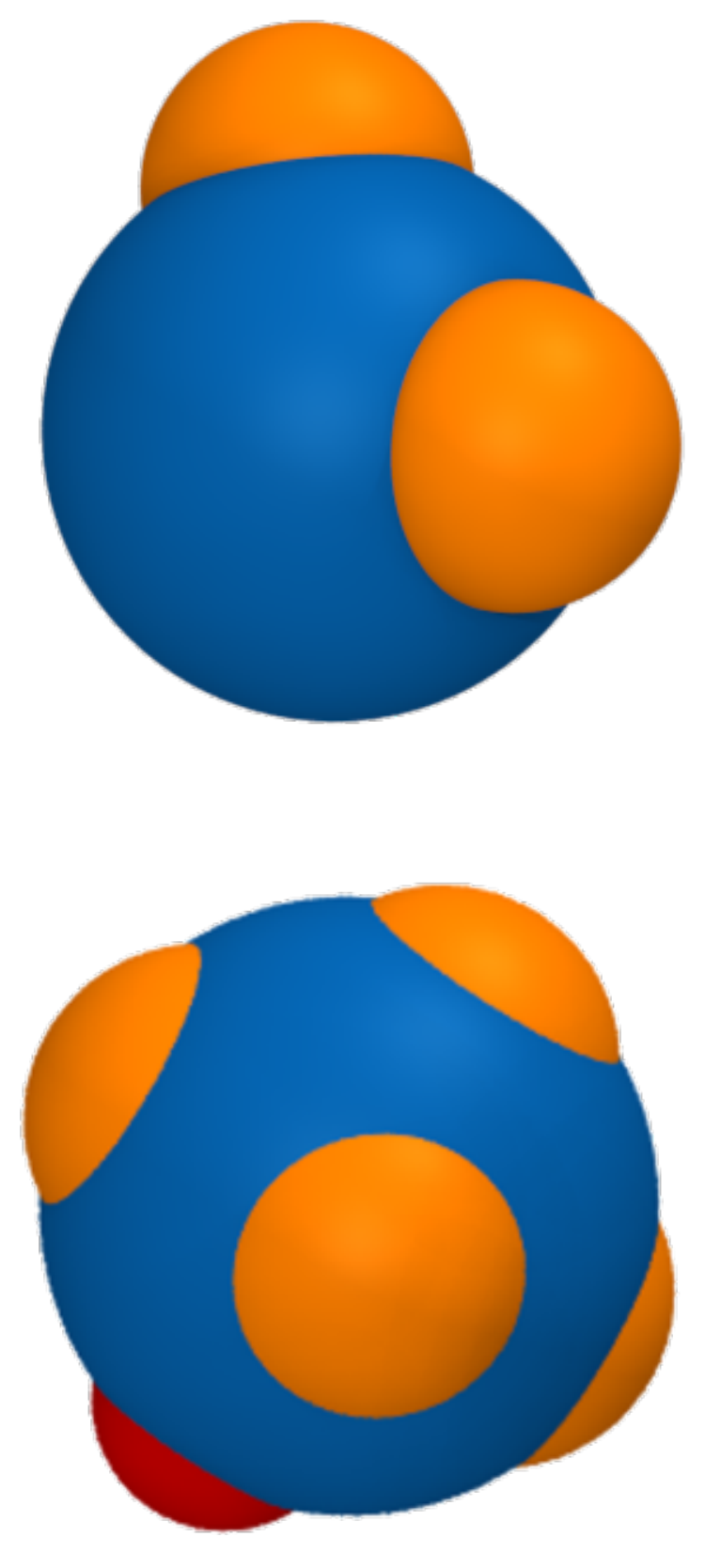}

\caption{Illustration of the patchy particle potential. \textbf{a.} This plot
shows the patchy particle potential between two particles with a diameter
of one, each with one patch. The potential is plotted as a function
of the relative distance between the two particles for orientations
corresponding to aligned or misaligned patches. If aligned, we observe
a stable minimum in the potential corresponding to particles binding.
If misaligned, there is no stable minimum, and the isotropic repulsion
prevents overlapping. See \citep{vijaykumar2017multiscale} for the
specific form of the potential. \textbf{b. }Two examples of angular
potentials used in this work, corresponding to one and two metastable
orientations. As the bindings between patches already fix two orientational
degrees of freedom, we only require a one dimensional angular potential
to completely fix the orientation. \textbf{c.} Examples of patchy
particles with two and six patches. The patches can also be of different
types corresponding to different interaction potentials, and they
can be turned on an off depending on the current conformation.}

\label{fig:patchyparts}
\end{figure*}

To validate MSM/RD schemes, we require an inexpensive model of molecules
capable of representing complex behavior observed in realistic MD
systems such as: translational and rotational diffusion, position
and orientation dependent pair interactions, orientation dependent
binding, multiple binding sites and conformation switching. We can
construct such a model based on patchy particles \citep{klein2014studying,newton2015rotational,schluttig2008dynamics}.
We model molecules as diffusive spherical particles with an isotropic
repulsive potential $U_{\text{isotropic}}$ to avoid overlapping;
an attractive isotropic part can also be incorporated. Patches are
then placed on the surface of the particles, and each patch produces
a short-range attractive potential with patches from other particles,
generating translational and rotational motion \citep{vijaykumar2017multiscale}.
The potential energy between patch $i$ of particle $A$ and patch
$j$ of particle $B$ can be decomposed into two parts. The first
part $U_{r}^{ij}$ depends only on the relative distance between the
patches, $r_{ij}$, and the types of the patches. It corresponds to
an attractive force that pulls patches together . The second part
$U_{\theta}^{ij}$ depends on the relative orientation, $\theta_{AB}$,
of the particles, and it is activated if two patches are close enough
to each other. This will favor specific relative orientations for
the different bindings between patches. In all the models used for
this work, the overall interaction potential between two particles,
$A$ and $B$, can be written in the following form:
\[
U_{AB}=U_{\text{isotropic}}(r_{AB})+\sum_{i,j=1}^{N_{A},N_{B}}\left(U_{r}^{ij}(r_{ij})+U_{\theta}^{ij}(r_{ij},\theta_{AB})\right),
\]
where $i$ runs over the patches of particle $A$ and $j$ runs over
the patches of particle $B$. $N_{A}$ and $N_{B}$ are the total
number of patches of $A$ and $B$ respectively. In general, particle
$A$ and $B$ can both have conformational changes; each combination
of conformations is allowed to have a completely different potential
energy. In this work, conformation changes will correspond to turning
on and off specific patches. Figure \ref{fig:patchyparts} shows the
potential between a pair of patchy particles with one patch, as well
as examples of orientation dependent potentials.

The position of the particles is simply given by the coordinates of
the center of the sphere $r(t)$, and their orientation $\theta(t)$
is given in terms of quaternions. In order to model the translational
and orientational diffusion of one particle, we use overdamped Langevin
dynamics. We assume the translational and rotational diffusion are
independent and both isotropic, so we obtain

\begin{align}
\frac{dr(t,\eta)}{dt} & =\frac{1}{\gamma}F(r,\eta)+\sqrt{2D(\eta)}\xi(t),\nonumber \\
\frac{d\Phi(\eta,t)}{dt} & =\frac{1}{\gamma_{\text{rot}}}\mathrm{T}(\theta,\eta)+\sqrt{2D^{\text{rot}}(\eta)}\xi_{\text{rot}}(t)
\end{align}

where $\Phi$ is the orientation in the axis-angle representation,
$\gamma$ and $\gamma_{\text{rot}}$ are the translational and rotational
damping coefficients; $F$ and $\mathrm{T}$ the force and torque
due to pair-interactions and external fields; $\eta$ is the conformation
of the particle; and $\xi(t)$ and $\xi_{\text{rot}}(t)$ each correspond
to three-dimensional Gaussian white noise. The force can be rewritten
in terms of the potential as $F=-\nabla U$; the torque is convenient
to leave explicitly since it is not trivial to write a potential in
terms of axis-angle variables or quaternions. These two equations
can be discretized using the Euler-Maruyama scheme \citep{higham2001algorithmic}
using a time-step $\delta t$\textcolor{blue}{.}
\begin{align}
r(t+\delta t,\eta) & =r(t,\eta)-\frac{\delta t}{\gamma}\nabla U(r,\eta)+\sqrt{2D(\eta)\delta t}\mathcal{N}(0,1),\nonumber \\
d\Phi(t,\eta) & =\frac{\delta t}{\gamma_{\text{rot}}}\mathrm{T}(\theta,\eta)+\sqrt{2D^{\text{rot}}(\eta)\delta t}\mathcal{N}(0,1).\label{eq:euler-maruyama}
\end{align}

The rotation represented by the change in axis-angle $d\Phi(t)$ can
be rewritten as a quaternion $d\theta(t)$ using Eq. \ref{eq:axangle2quaternion}.
The new orientation is simply given by the quaternion product $\theta(t+\delta t)=d\theta(t)\theta(t).$
In each case, $\mathcal{N}(0,1)$ represents an independent three-dimensional
vector with each entry a normal random variable with mean zero and
variance $1$. As the diffusion coefficients depend on the conformation,
it is convenient to assume the switching of conformation $\eta$ is
modeled with an MSM using a fixed lag-time $\tau$ that is a multiple
of $\tau=n\delta t$, $n$ a positive integer. This is not strictly
required, but it simplifies the implementation since the conformation
change occurs always at the end of a time step. The forces and torques
are calculated directly from potential energies like the ones shown
in Fig. \ref{fig:patchyparts}.

This model satisfies all the requirements we established at the beginning
of this section. It can be generalized to non-isotropic and coupled
rotational and translational dynamics \citep{delong2015brownian,schluttig2008dynamics},
and molecules can even be modeled by multiple overlapping beads with
reaction patches \citep{schluttig2008dynamics}.

\section*{IV. Results}

\begin{SCfigure*}
\begin{minipage}[t]{0.72\textwidth}%
\textbf{$\centering$ a.\includegraphics[width=0.66\textwidth]{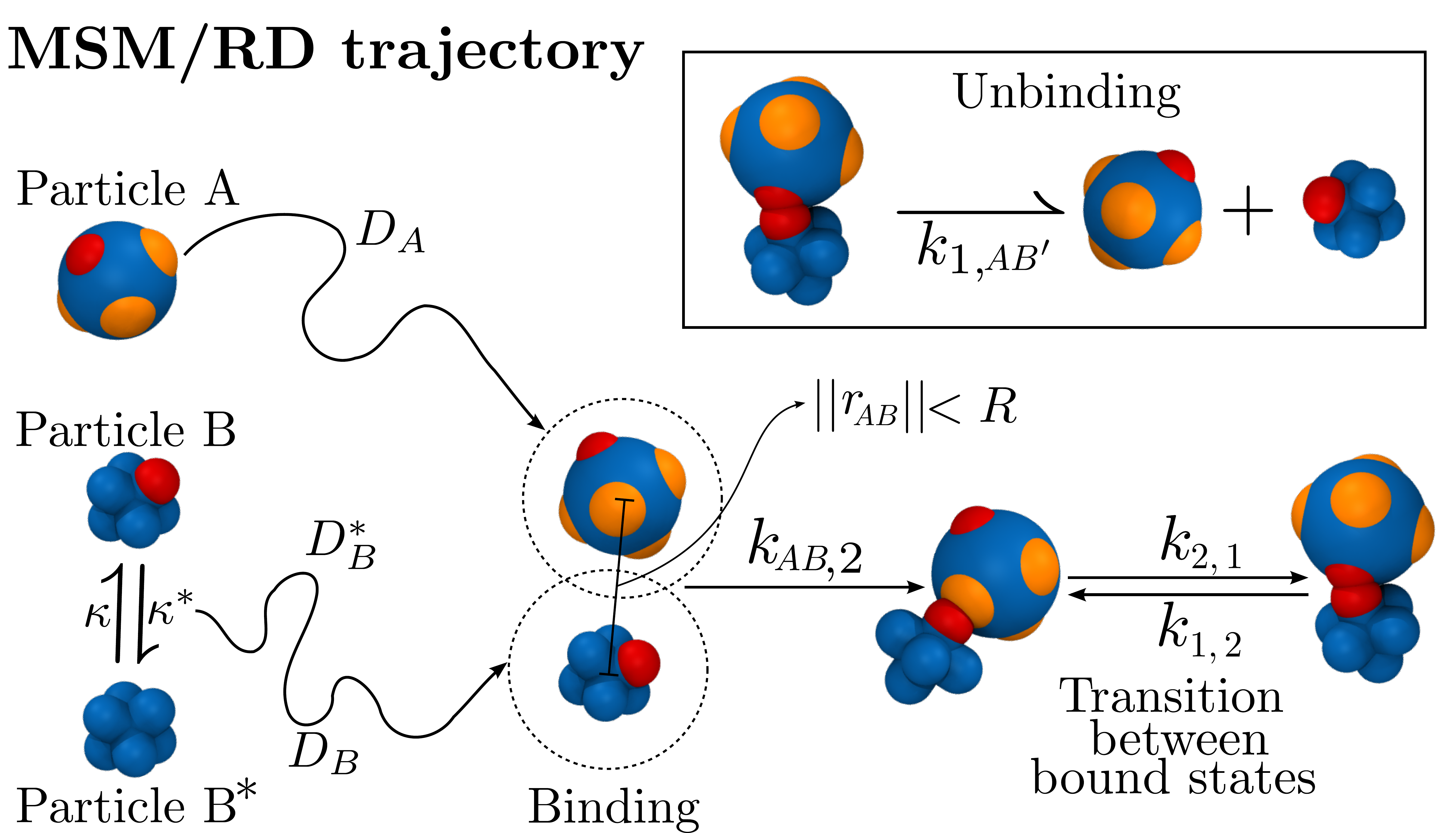}b.}\includegraphics[width=0.25\textwidth]{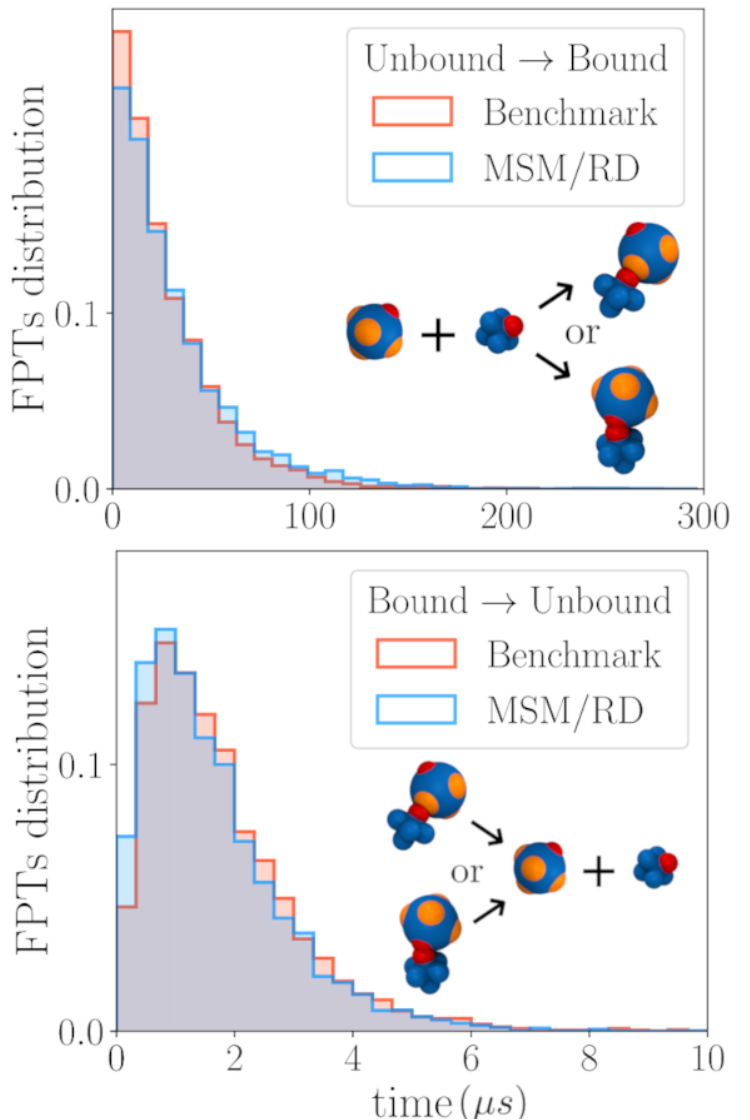}

\medskip{}

\textbf{$\centering$ c.}\includegraphics[width=0.95\textwidth]{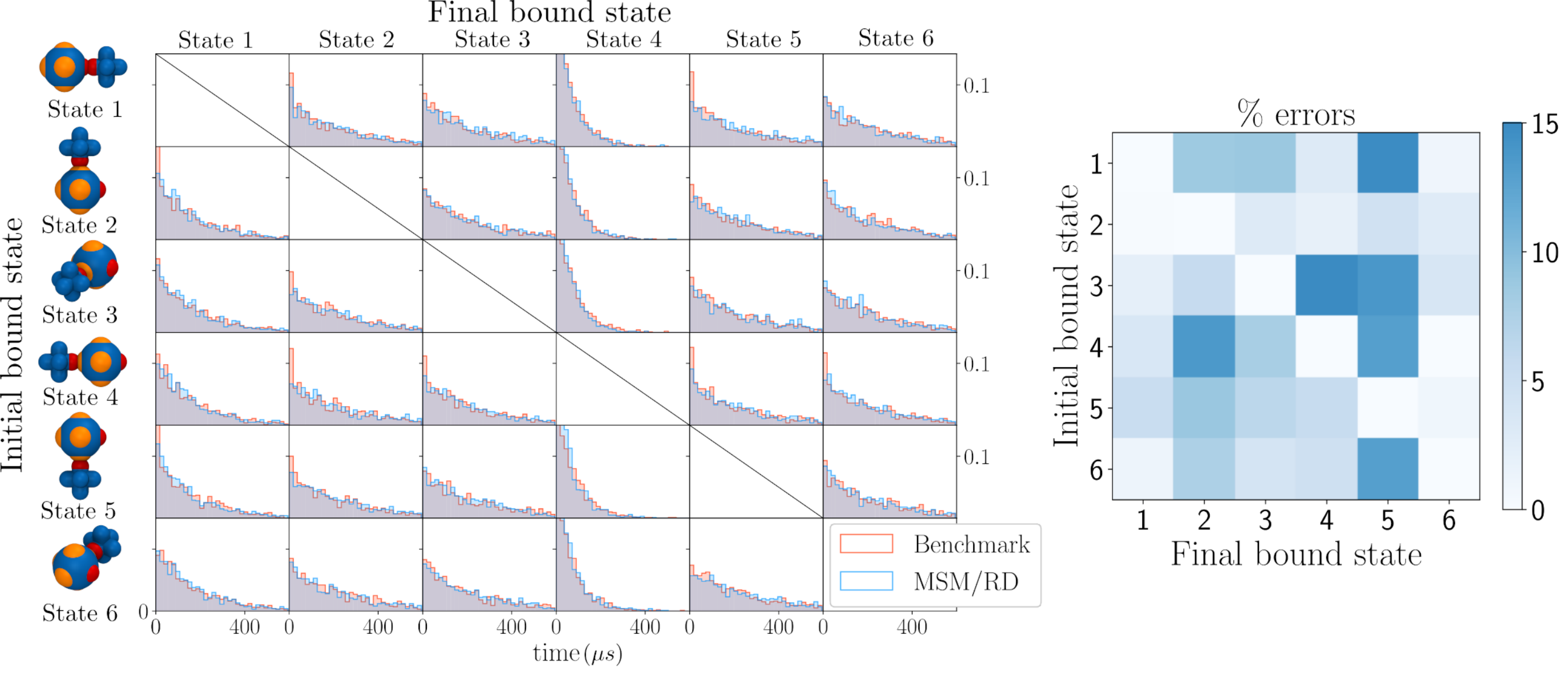}%
\end{minipage}

\caption{Illustrations and results of the MSM/RD scheme for the protein-protein
system. \textbf{a.} MSM/RD sample trajectory. Particle $A$ diffuses
with coefficient $D_{A}$, and particle $B$ with coefficient $D_{B}$
or $D_{B}^{*}$ depending on its conformation. If the relative distance
satisfies $\left\Vert r_{AB}\right\Vert <R$, we switch from the non-interacting
to the transition regime. Here, the particles can transition to one
of the six bound states with rate $k_{AB,n}$ that depends on their
relative configuration $x_{AB}=(r_{AB},\theta_{AB})$ and the final
bound state $n=1,\dots6$. From a bound state $n$, the compound can
transition to another bound state $m$ with rate $\kappa_{n,m},$
or it can unbind to another relative configuration $x_{AB}^{'}$ with
rate $k_{n,AB^{'}}$. MSM/RD provides a piecewise constant approximation
of all the configuration-dependent rates. \textbf{b.} Comparisons
of first passage time (FPT) distributions between the benchmark and
MSM/RD from an unbound state to any bound state and vice versa. Each
distribution was calculated with $5000$ simulations. \textbf{c.}
Comparison of FPT distributions for all transitions between the six
possible bound states, each calculated over $1000$ simulations. The
blue grid shows the relative error of the corresponding rates, $\kappa_{n,m}$,
calculated as the inverse MFPT. The average percentage error is of
$5\%,$while the maximum is of $16\%$.}

\label{fig:patchyProteinResults}
\end{SCfigure*}

\begin{figure*}[t]
$\qquad$%
\begin{minipage}[t]{0.9\textwidth}%
\centering\textbf{a.}\includegraphics[width=0.17\textwidth]{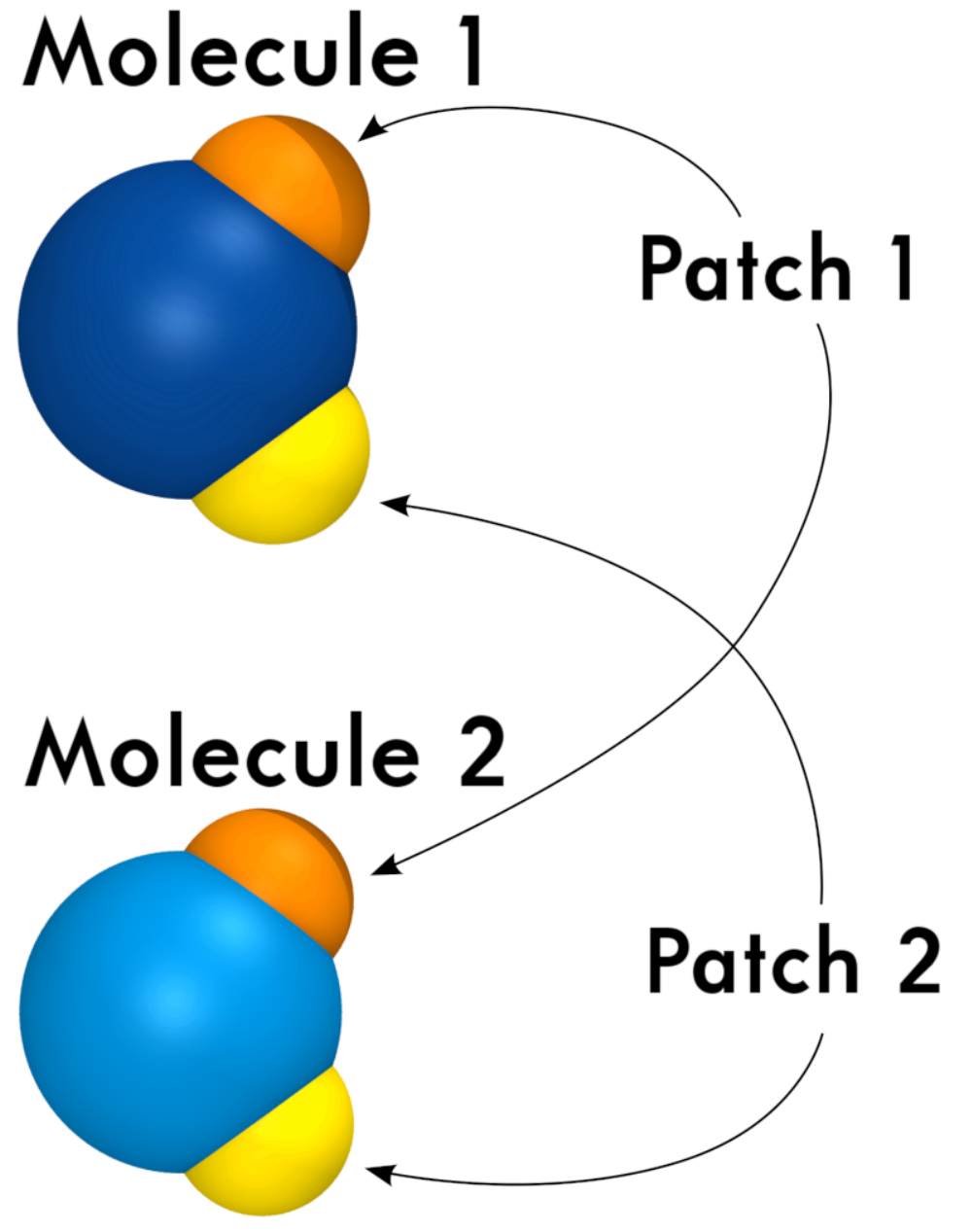}\includegraphics[width=0.45\textwidth]{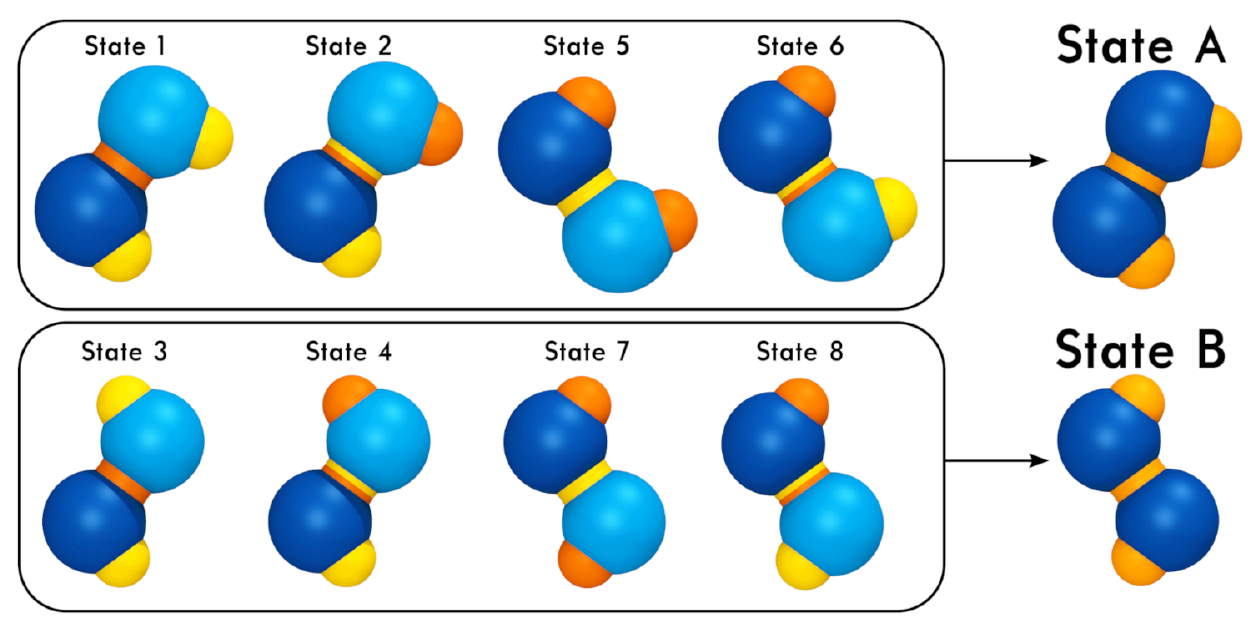}\textbf{b.}\includegraphics[width=0.33\textwidth]{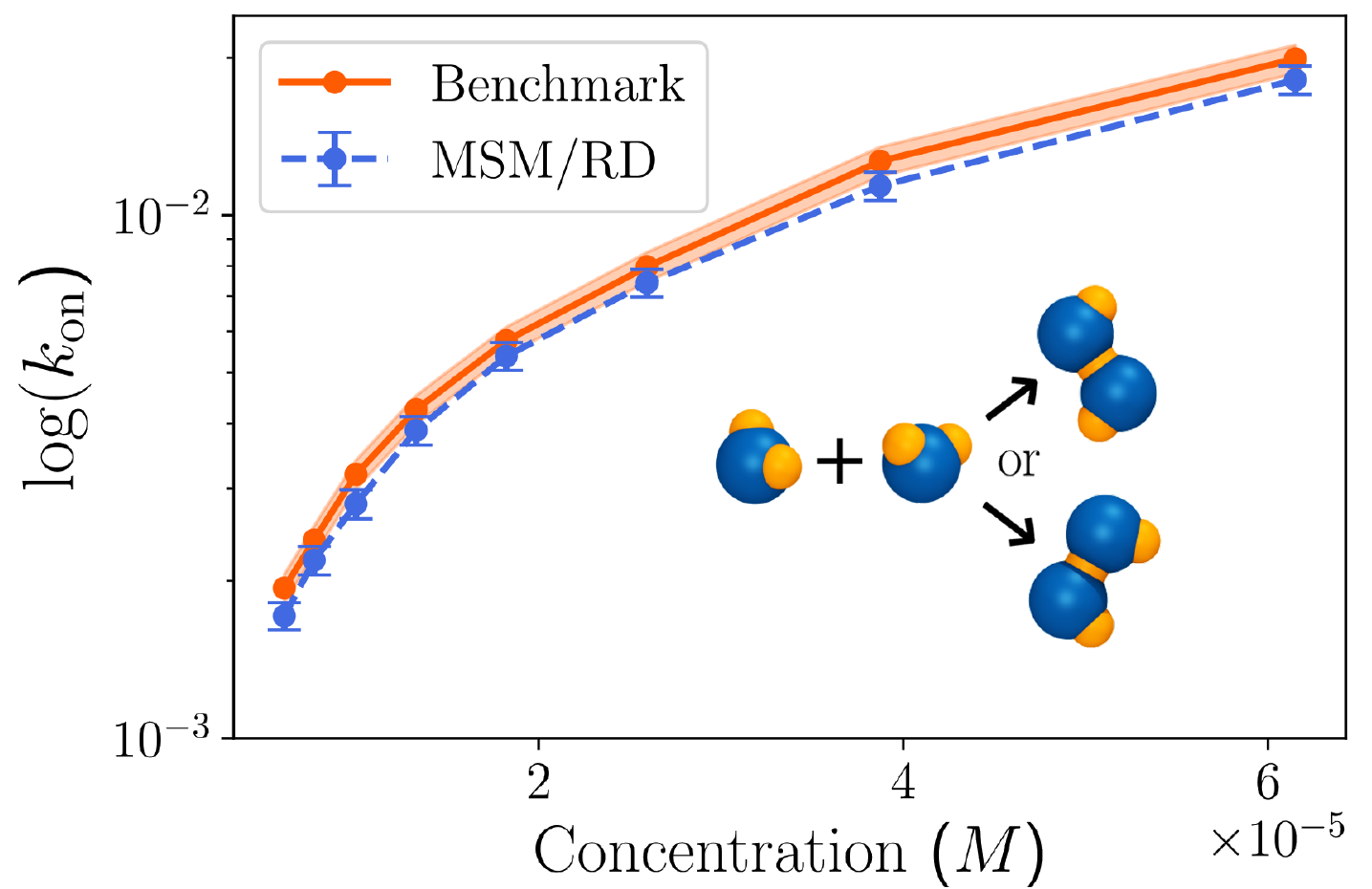}

\textbf{c.}\includegraphics[width=0.33\textwidth]{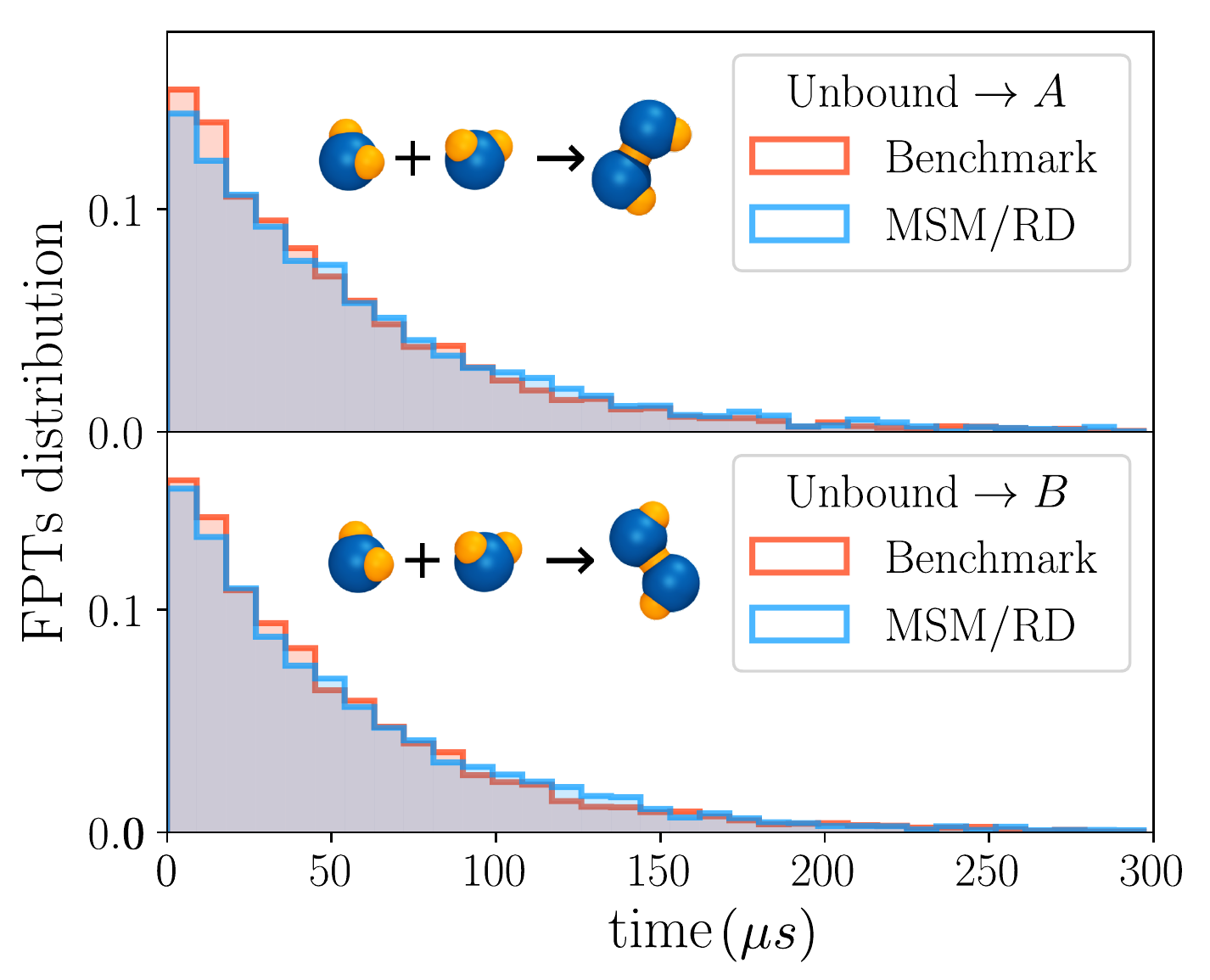}\includegraphics[width=0.33\textwidth]{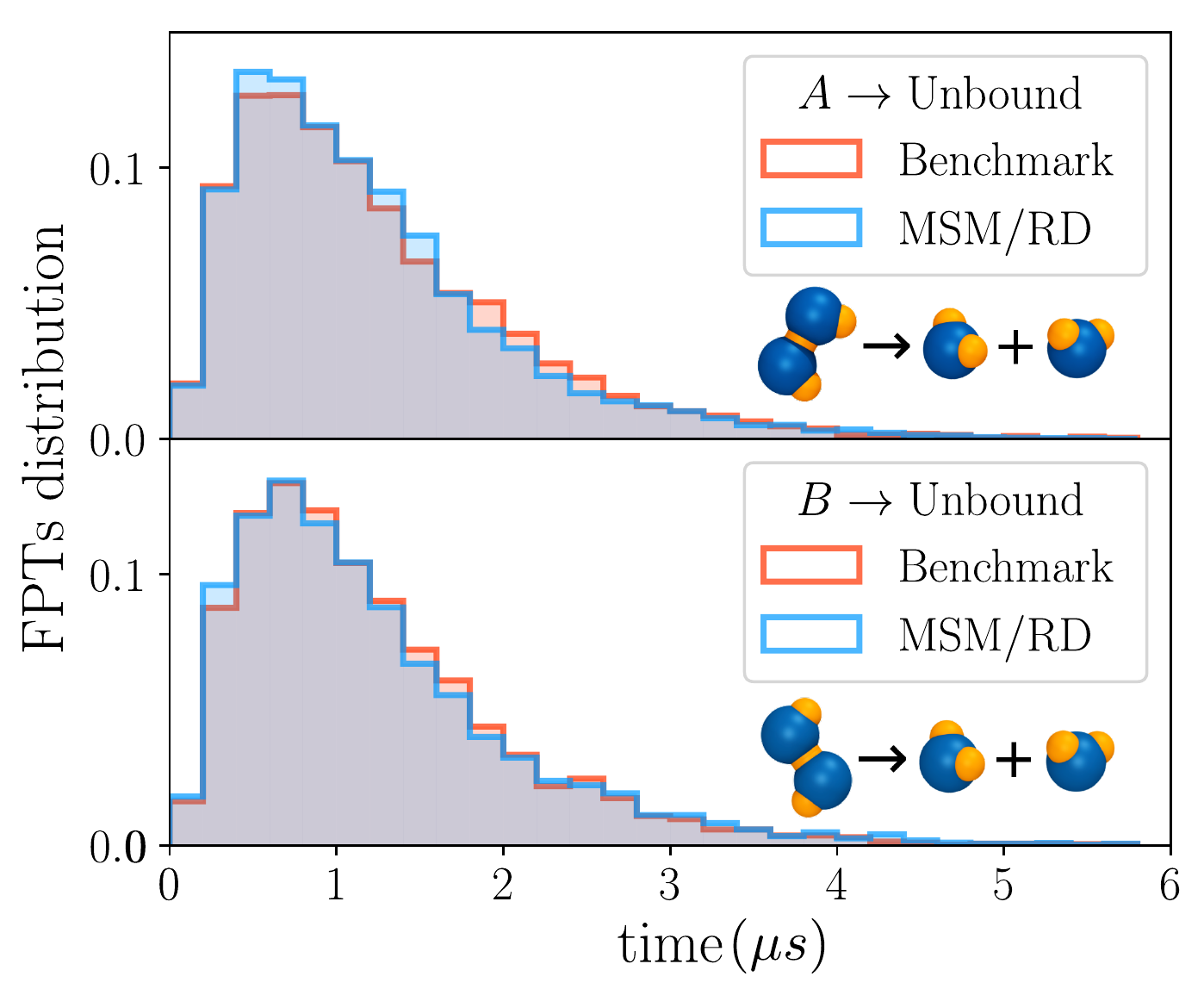}\includegraphics[width=0.33\textwidth]{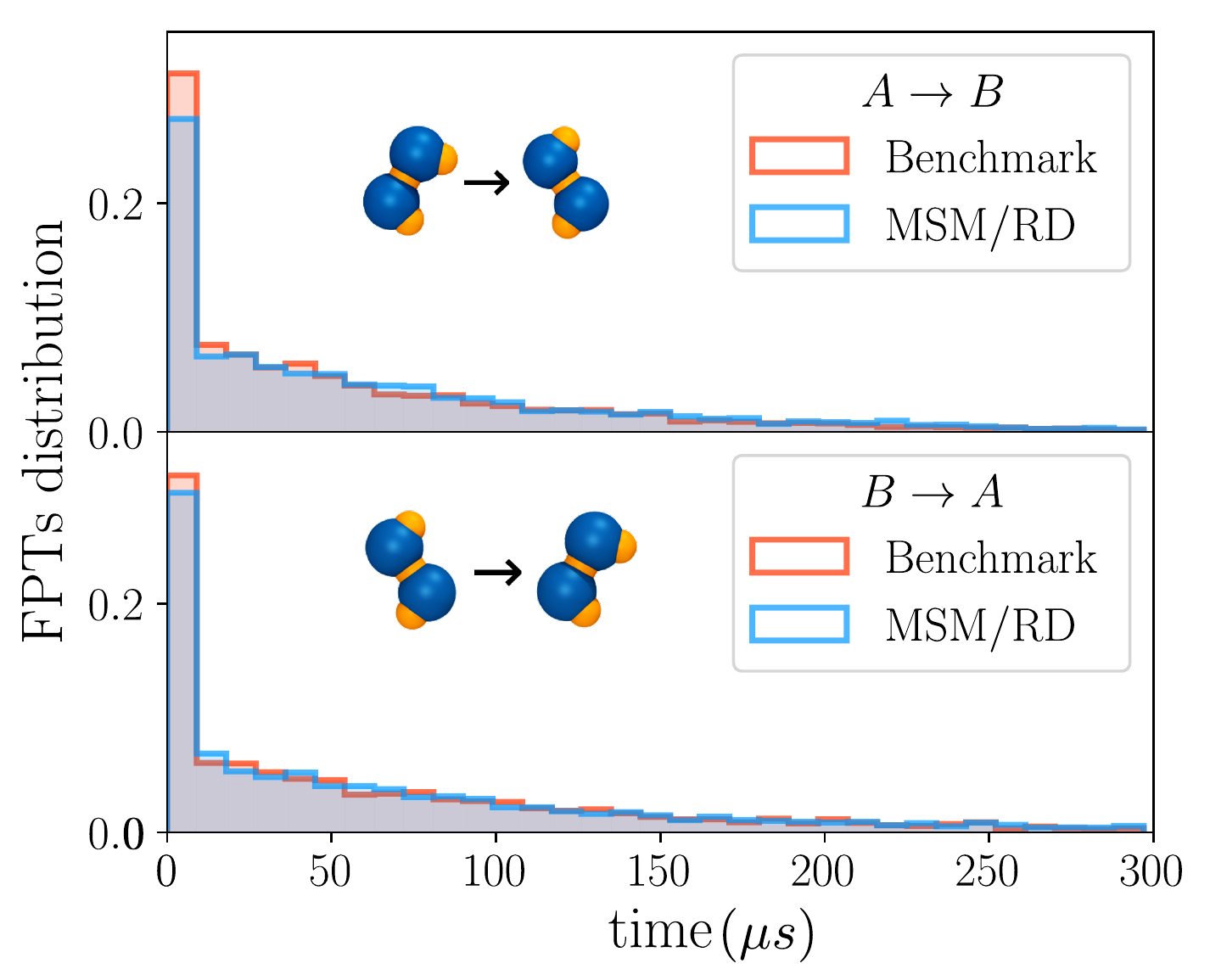}%
\end{minipage}

\caption{Illustrations and results of the MSM/RD implementation for two identical
interacting molecules with two interacting patches each and two stable
angular configurations. $\mathbf{\mathbf{a}.}$ The two molecules
can bind in eight different ways (eight bound states). For illustration
purposes, particle one is shown in dark blue and particle 2 in light
blue; the first patch is shown in orange and the second one in yellow.
All these states collapse into two bound states: $A$ and $B$. $\mathbf{\mathbf{b}.}$
Comparison of the on-rate, transition from unbound to either A or
B state, for different molar concentrations. Each point was calculated
as the inverse of the MFPT obtained from $5000$ simulations; the
error bars represent the standard deviation over $2000$ bootstrapped
samples. Note that in the generation of the MSM faster time-scales
are neglected; therefore, it is expected that MSM/RD produces slightly
slower results than the benchmark. \textbf{c. }Comparisons of the
FPT distributions obtained with MSM/RD and the benchmark for six cases:
from the unbound state to the two bound states and vice versa, and
between the bound states. Each distribution was computed using 5000
simulations. These are shown next to each graph and they are all in
$\mu$s. Note in the last two histograms there is a time-scale separation.
This corresponds to the difference between direct transitions between
the bound states and transitions that first unbound and later rebound
in a different bound state. }

\label{fig:dimerallstates}
\end{figure*}
\begin{SCfigure*}
\textbf{$\centering$ \includegraphics[width=0.74\textwidth]{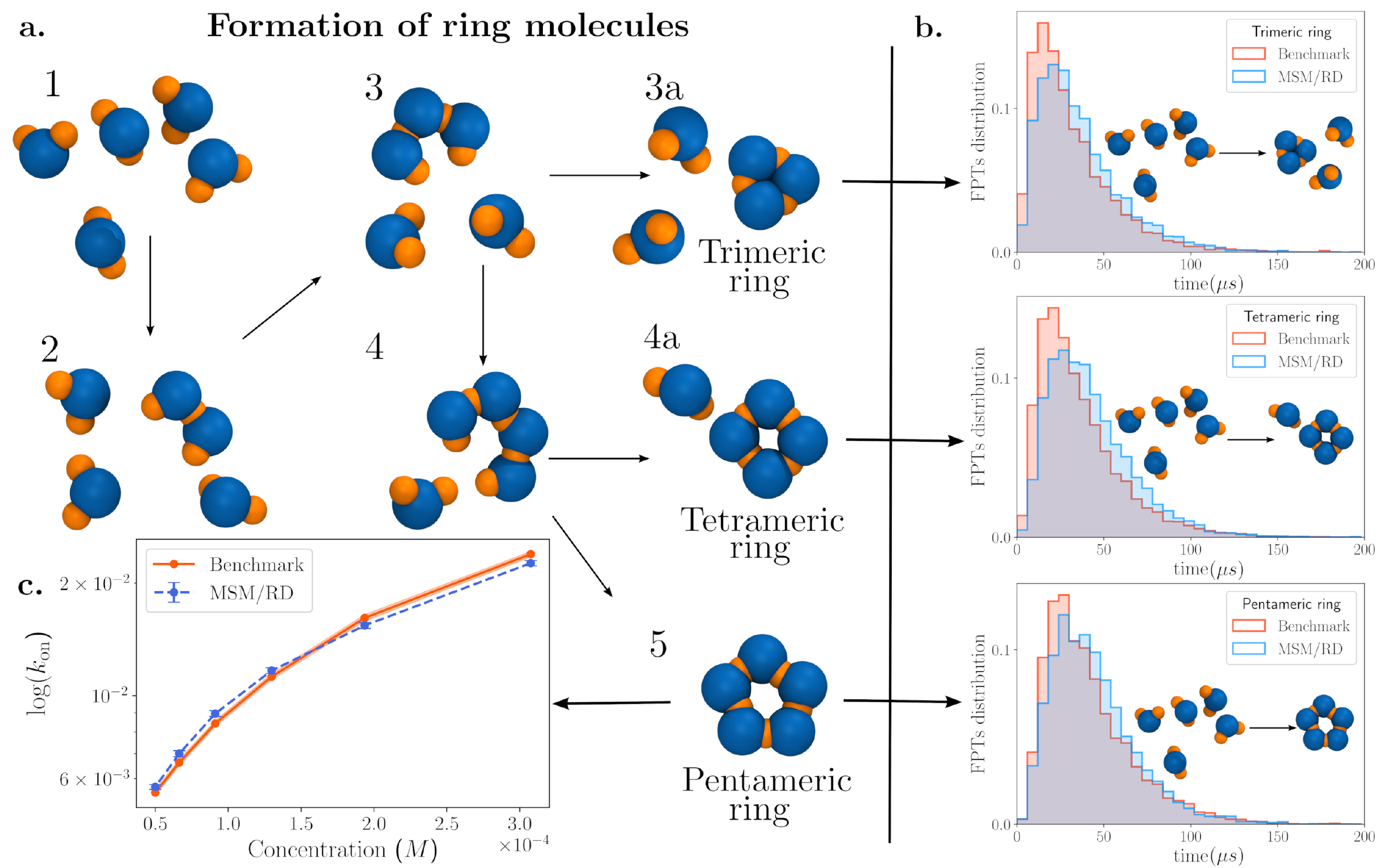}}

\caption{Illustrations and multiparticle MSM/RD results for the formation of
ring molecules. \textbf{a. }Diagram showing the formation of the trimeric,
tetrameric and pentameric ring molecules. \textbf{b. }Comparisons
of the FPT distributions obtained with multiparticle MSM/RD and the
benchmark for the formation of the trimeric, tetrameric and pentemric
ring molecules. The results were obtained from $5000$ simulations
for each case. \textbf{c. }Comparison of the rate at which a pentameric
ring is generated for different molar concentrations. Each point was
calculated as the inverse of the MFPT obtained from $1000$ simulations;
the error bars represent the standard deviation over $500$ bootstrapped
samples.}

\label{fig:ringmol-formation}
\end{SCfigure*}

To test and verify MSM/RD, we construct an MD benchmark model of molecules
that is simple enough such that we can produce a large amount of data,
but complex enough so it models complex behavior observed in realistic
MD systems. This model is based on patchy particles \citep{klein2014studying,newton2015rotational,schluttig2008dynamics};
we model molecules as spherical particles with isotropic diffusion
and an isotropic repulsive potential to avoid overlapping. Patches
are then placed on the surface of the particles, and each patch produces
a short-range configuration-dependent attractive potential with patches
from other particles, generating translational and rotational motion,
see the methods section for details. This model is the basis for all
the benchmark models in this section.

In the following examples, it is not necessary to parametrize the
diffusion operator since it is the same for both the benchmark and
the MSM/RD simulation. This serves to isolate the dynamics of $\mathbb{Q}(x)$
and validate the coupling mechanism. For general protein-protein systems,
one needs to extract the diffusion coefficients/matrices from the
MD data using well documented methods \citep{bullerjahn2020optimal,linke2018fully}
(SI Appendix \ref{app:diffusionEstimation}).

\subsection*{A. MSM/RD for protein-protein systems \label{subsec:protein-proteinMSMRD}}

The benchmark model consists of two molecules, $A$ and $B$ (Fig.
\ref{fig:patchyProteinResults}a), represented by different patchy
particles. Particle $A$ has only one conformation and six binding
patches: five of them have the same attraction potential (yellow),
and the other one has a stronger attraction potential (red). Particle
$B$ has two conformations. In one conformation ($B$), it has one
binding patch (red), and in the other one ($B^{*})$, the patch is
turned off, and it cannot bind. Each binding allows only one meta-stable
relative orientation, yielding a total of six possible bound states.
The diffusion of $B$ depends on its conformation, and it is visualized
as a three-dimensional asterisk to distinguish its orientation. We
illustrate an MSM/RD trajectory of the system on Fig. \ref{fig:patchyProteinResults}a.

To parametrize the MSM/RD scheme, we simulate the benchmark MD model
with specific settings to mimic a common MD simulation. We assume
both molecules have a diameter of $5\,nm$, which is a typical size
for a real protein. We simulate using periodic boundary conditions
and a cube with edge-length of $25\text{nm}$ as unit cell. Each simulation
runs for $6\times10^{6}$ time steps of $1\text{\ensuremath{\times10^{-5}}}\mu s$
each, yielding a total simulation time of $60\mu s$. We run $600$
of these simulations independently, and we use them to parametrize
the MSM/RD scheme following the steps illustrated in SI Appendix \ref{app:paramMSM/RD}.

In Fig. \ref{fig:patchyProteinResults}, we compare the MSM/RD results
against the MD benchmark. We calculate the first passage times (FPTs)
of a given transition by running the benchmark and the MSM/RD simulations
in equal conditions, and we run the same number of FPT samples for
each model. Figure \ref{fig:patchyProteinResults}b compares the FPTs
distribution from the unbound state to any bound states and vice versa.
The left panel of Fig. \ref{fig:patchyProteinResults}c compares the
FPT distributions for all the possible transitions between bound states.
Note these transitions include pathways that start at a bound state,
unbind completely and end in another bound state, so it is ideal to
evaluate if the MSM/RD produces an accurate coupling. The right panel
of Fig. \ref{fig:patchyProteinResults}c shows the MSM/RD scheme percent
error for the transition rates between bound states. Overall, MSM/RD
can reproduce the dynamics of the MD benchmark with good accuracy.

\subsection*{B. MSM/RD for dimer of two-patch particle \label{subsec:implementation1}}

The benchmark model consists of two identical molecules, each with
two equally strong binding patches. The molecules can bind together
through any of their two binding sites. Unlike the previous example,
we allow for two meta-stable relative orientations per patch binding,
allowing for a conformation change in the bound configuration. 

As the molecules have two patches each, they can bind in four different
ways; each of these has two stable relative orientations, so this
system has a total of eight meta-stable bound states. However, as
they are all identical, many of these eight states are indistinguishable
from each other and can all be collapsed into two functional states,
$A$ and $B$. This is depicted graphically in Fig. \ref{fig:dimerallstates}a.
Nonetheless, note each of these meta-stable states corresponds to
a different relative position and orientation between the molecules.
We parametrize the MSM/RD scheme with the same setup as in the protein-protein
system example, see SI Appendix \ref{app:paramMSM/RD}.

In Fig. \ref{fig:dimerallstates}, we compare MSM/RD results against
the MD benchmark. We calculate FPTs for both the MD benchmark and
the MSM/RD simulations in equal conditions. In Fig. \ref{fig:dimerallstates}b,
we show the binding rates as a function of concentration, where each
binding rate is calculated as the inverse of the mean first passage
time (MFPT). The concentration is adjusted by changing the edge-length
of the simulation box, starting at $30\,nm$ and increasing $5\,nm$
for each point. In Fig. \ref{fig:dimerallstates}c, we compare the
FPTs distributions for several relevant transitions. Even when the
original simulations to parametrize the scheme ran for only $30\,\mu s,$
the MSM/RD scheme produces excellent results for transitions with
higher MFPTs.

\subsection*{C. Multiparticle MSM/RD: formation of pentameric ring \label{sec:multiparticle}}

\begin{table}
\centering

\caption{Comparison between the MD benchmark and MSM/RD of the rates of formation
of pentameric rings for different concentrations. The rates were calculated
as the inverse of the MFPTs averaged over 1000 simulations; the uncertainties
represent the standard deviation over 100 bootstrapped samples.}
\begin{tabular}{lcccc}
Rate & Concentration & MD Benchmark & MSM/RD & \% Error\tabularnewline
\midrule
 & $3.08\cdot10^{-4}M$ & {\small{}$23.96\pm0.44$} & {\small{}$22.67\pm0.41$} & {\small{}5.4\%}\tabularnewline
 & $1.94\cdot10^{-4}M$ & {\small{}$16.16\pm0.33$} & {\small{}$15.43\pm0.31$} & {\small{}4.5\%}\tabularnewline
{\small{}$k_{\text{on}}$} & $1.30\cdot10^{-4}M$ & {\small{}$11.27\pm0.23$} & {\small{}$11.72\pm0.21$} & {\small{}3.4\%}\tabularnewline
{\small{}($\text{ms}^{-1}$)} & $9.11\cdot10^{-5}M$ & {\small{}$8.42\pm0.15$} & {\small{}$8.95\pm0.16$} & {\small{}6.3\%}\tabularnewline
 & $6.64\cdot10^{-5}M$ & {\small{}$6.62\pm0.13$} & {\small{}$6.99\pm0.13$} & {\small{}5.7\%}\tabularnewline
 & $4.99\cdot10^{-5}M$ & {\small{}$5.50\pm0.09$} & {\small{}$5.57\pm0.08$} & {\small{}3.6\%}\tabularnewline
\end{tabular}

\label{tab:ratesRings}
\end{table}

We develop and implement the first multiparticle MSM/RD scheme to
study the formation of pentameric ring molecules (inspired by \citep{klein2014studying}).
The benchmark MD model is a modified version of the two-patch dimer
model. It consists again of two identical molecules, each with two
equally strong binding sites. Unlike the previous example, we only
allow one meta-stable relative orientation per patch binding. We further
increase the binding strength such that unbinding events are very
rare and not observed in the timescales of interest. The particles
can bind with each other forming chains, which eventually can close
forming either trimeric, tetrameric or pentameric ring structures
(Fig. \ref{fig:ringmol-formation}a).

We parametrize the MSM/RD scheme with the same setup as in the protein-protein
system example, see SI Appendix \ref{app:paramMSM/RD}. The multiparticle
MSM/RD scheme requires modifications to the two-particle MSM/RD algorithm.
These modifications are shown in SI Appendix \ref{app:MSM/RD-schemes}.
We further need to estimate the diffusion coefficients of the multiparticle
chains. We employ standard methods to estimate them (SI Appendix \ref{app:diffusionEstimation}).

In Fig. \ref{fig:ringmol-formation}b, we compare MSM/RD results against
the MD benchmark. We calculate FPTs for the formation of all the three
ring molecules, using both the MD benchmark and the MSM/RD simulations
in equal conditions: five particles with random positions and orientations
placed in a simulation box of edge-length of $30$nm with periodic
boundaries. In Fig. \ref{fig:dimerallstates}c, we show the rate of
formation of pentameric rings for different concentrations by changing
the simulation box size. In Table \ref{tab:ratesRings}, we show the
exact values and relative errors of the rates plotted in Fig. \ref{fig:ringmol-formation}c.

Note MSM/RD is not as good at approximating the formation of trimeric
and tetrameric rings (Fig. \ref{fig:ringmol-formation}b). This is
due to MSM/RD modeling the particle-chains in steps 3 and 4 of Fig
\ref{fig:ringmol-formation}a as a fixed structure, while in the MD
benchmark the chain is flexible, allowing for patches to get closer
together, which increases the rate at which the ring is closed. This
could be fixed by using a new MSM to describe the dynamics between
the three or four particle chains and an additional particle. Nonetheless,
note the rates of formation of pentameric rings are not affected by
these problem since they are conditioned on not having trimeric or
tetrameric rings forming beforehand. In its current formulation, MSM/RD
multiparticle implementations are limited to non-crowded environments
since only pair interactions are parametrized. It is important to
take these issues into account when implementing MSM/RD applications.

\section*{V. Discussion}

We presented a coarse-grained model of molecular kinetics based on
hybrid switching diffusions. With this model, we developed a robust
framework for coupling Markov models of molecular kinetics with particle-based
reaction diffusion (MSM/RD). Based on this framework, we derived one
possible MSM/RD scheme by discretizing the underlying equation (Eq.
\ref{eq:diffMS_MSMRD}), generalizing previous approaches \citep{dibak2018msm}.
We implemented and verified it for three benchmark systems: the first
two involve two protein-protein systems, while the third one is a
multiparticle system to model the formation of pentameric molecules.
We obtained an excellent agreement between the FPT distributions and
reaction rates of relevant transitions.

The framework is well-suited to model protein-ligand binding in large
domains and time-scales as in the previous work \citep{dibak2018msm}.
Given enough data for the parametrization, it is also suited to model
protein-protein dynamics since it incorporates arbitrary orientations,
conformation switching and multiple binding sites. To parametrize
the MSM/RD scheme for protein-protein systems, we would need the MD
data of the two proteins interacting and individually, both in small
simulation boxes. The interacting proteins data would serve to parametrize
the scheme in the bound and transition regime, similarly to the works
\citep{BuchFabritiis_PNAS11_Binding,SilvaHuang_PlosCB_LaoBinding,PlattnerNoe_NatComm15_TrypsinPlasticity,PlattnerEtAl_NatChem17_BarBar}
with the addition of the transition states, which might require a
slightly larger box. The individual molecules data would serve to
parametrize the scheme in the non-interacting regime. The resulting
MSM/RD scheme could run simulations at much larger time- and length-scales
than those allowed by MD.

The multiparticle implementation of MSM/RD has promising applications
to the study of self-assembly of structures composed of several copies
of the same molecule (or a small set of molecules), such as virus
capsids \citep{arkhipov2006stability} or soft matter self-assembly.
This setting is ideal since we only need MD data of one pair (or a
few pairs) of molecules in a small simulation box to parametrize an
MSM/RD multiparticle simulation, which could potentially model the
formation of the full capsid.

The main caveat of MSM/RD is that the parametrization requires a large
amount of MD data, which is not yet possible to obtain for most systems
of interest. However, given the increasing computational power, more
and more systems will soon be within reach of MSM/RD. The scheme might
also become less effective in the presence of long-ranged interactions,
though it might be possible to incorporate them into the dynamics
of the non-interacting-regime using coarse-grained potentials \citep{davtyan2012awsem,wang2021multi}.
Finally, in its current form, the MSM/RD multiparticle implementation
only takes into account pair interactions, and thus the scheme is
not yet adequate for crowded multi-molecular environments.

Although application-dependent, one can expect MSM/RD to reduce computational
cost by several orders of magnitude in comparison to MD. MD simulations
propagate the position and velocity of every atom, which corresponds
to several thousands of degrees of freedom in an average protein-protein
simulation. MSM/RD only propagates two independent Brownian bodies
together with an MSM. This corresponds to at most 14 degrees of freedom,
6 for the position/orientation and one for the MSM (per molecule).
Finally, considering that MSM/RD can operate in larger domains with
larger time steps, equivalent MD simulations would need to increase
dramatically the number of solvent molecules yielding an exploding
number of degrees of freedom, while still limited to small time steps.

\subsection*{Software}

To enable reproducibility and implementation of this work, we developed
the MSM/RD software package, a C++/python package. All the code and
software developed for this work are open source and available under
an MIT license in \href{https://github.com/markovmodel/msmrd2}{github.com/markovmodel/msmrd2}
and Zenodo \citep{delRazoMSMRD}. The data used in this work was produced
using the MSM/RD software.

\acknow{We acknowledge support by the European Commission (ERC CoG 772230),
German Ministry for Education and Research (Berlin Institute for the
Foundations of Learning and Data BIFOLD), Deutsche Forschungsgemeinschaft
(SFB1114/C03, SFB1114/A04 and TRR186/A12), the Berlin Mathematics research center
Math+ (project AA1-6), and the Dutch Institute for Emergent Phenomena
at the University of Amsterdam. M.J.R. thanks Hong Qian for helpful
discussions over the course of this work. We also thank two anonymous
reviewers that greatly improved the presentation of this work.}

\showacknow{} 

\subsection*{Data Availability}

The data and scripts to produce the plots in this work are available
in Zenodo \citep{delRazoMSMRDdata}. The complete dataset that support
the findings of this study are available from the corresponding author
upon reasonable request.

\subsection*{References}

\begin{btSect}{msmrd-refs,own}
\btPrintCited
\end{btSect}

\appendix
\setcounter{figure}{0} \renewcommand{\thefigure}{S.\arabic{figure}}
\setcounter{algorithm}{0} \renewcommand{\thealgorithm}{S.\arabic{algorithm}}
\setcounter{table}{0} \renewcommand{\thetable}{S\arabic{table}}
\setcounter{equation}{0} \renewcommand{\theequation}{S\arabic{equation}}
\renewcommand{\partname}{}
\renewcommand{\thepart}{} 
\renewcommand{\thesection}{\Alph{section}} \setcounter{section}{0}
\renewcommand{\thesubsection}{\Alph{section}.\arabic{subsection}}
\renewcommand{\part}[1]{\twocolumn[\huge{\textbf{{#1}}}\vspace{10mm}]}

\begin{btUnit}

\part{Supplementary information: Multiscale molecular kinetics by coupling
Markov state models and reaction-diffusion dynamics. {\normalsize{}}\protect \\
{\normalsize{}Mauricio J. del Razo, Manuel Dibak, Christof Schütte
and Frank Noé}}

\section{Molecular kinetics as hybrid switching diffusion processes \label{app:fullTheory}}

\subsection{A first derivation of a hybrid switching diffusion process \label{app:CD1stderivation}}

We first introduce the hybrid switching diffusion model for a simple
scenario, which can be easily generalized to more complex cases. Consider
one macromolecule $A$ with $N$ atoms and assume the molecule follows
overdamped Langevin dynamics, the corresponding stochastic differential
equation (SDE) is 

\begin{align}
dq(t) & =-\nabla U(q)dt+\Sigma dW(t)\label{eq:overdampLan}
\end{align}
with $q(t)=[q_{1},\cdots,q_{N}]$ and $q_{i}$ the three-dimensional
positions of the i$^{\text{th}}$ atom, $U(q(t))$ the potential function
of the interactions, $\Sigma$ a matrix related to the diffusion tensor
and $W(t)$ is an $3N$ dimensional vector of standard Brownian motions.
This equation describes the trajectories of the molecule's atoms.
Alternatively, we can focus on the dynamics of the corresponding probability
density function. This is given in terms of the Fokker-Planck equation 

\begin{align}
\partial_{t}f(q,t) & =\mathcal{L}f(q,t)\nonumber \\
\mathrm{with}\qquad\mathcal{L}f(q) & =\sum_{i,j=1}^{d}\left[\partial_{q_{i}q_{j}}^{2}(D_{ij}f(q))\right]+\nabla U(q)\cdot\nabla f(q)+\label{eq:overLoperator}\\
 & \nabla^{2}U(q)f(q),\nonumber 
\end{align}
where $D_{ij}$ are the entries of the diffusion tensor $D=\frac{1}{2}\Sigma\Sigma^{\mathrm{T}}$.
The operator $\mathcal{L}$ is called the infinitesimal generator.
If $\Sigma$ is a diagonal matrix (uncorrelated Brownian motions)
with the same constant value $\sigma$ in each entry, we recover the
well-known Smoluchowski model $\mathcal{L}f=D\nabla^{2}f+\nabla\cdot(f\nabla U)$,
with $D=\sigma^{2}/2$. Consider the molecule as a rigid body with
its center of mass given by $r(t)$, and its orientation given by
the quaternion $\theta$. To better analyze the molecule's configurations,
we change to the frame of reference of the molecule:
\begin{gather}
\tilde{q}_{i}(t)=R_{\theta}^{-1}[q_{i}(t)-r(t)]\qquad\Rightarrow\qquad q_{i}(t)=r(t)+R_{\theta}\tilde{q}_{i}(t),\label{eq:frametransf}
\end{gather}
where $r(t)=\sum_{i=1}^{N}q_{i}(t)/N$; $R_{\theta}$ is a rotation
matrix corresponding to a rotation by the quaternion $\theta$; $R_{\theta}^{-1}$
is the inverse rotation matrix; and $\tilde{q}(t)=[\tilde{q}_{i}(t),\dots\tilde{q}_{N}(t)]$
are the coordinates of the atoms in the molecules's frame of reference.
As $r$ and $\theta$ are both functions of $q$, we could use Ito's
formula in conjunction with Eq. (\ref{eq:overdampLan}) to derive
SDEs for $r$ and $\theta$. We could also use the relations just
derived to obtain SDEs for the atom variables in the new coordinates
$\tilde{q}$. Although this can be written explicitly, it is more
convenient for our analysis to write directly the Fokker-Planck equation
in the new variables. As the Fokker-Planck equation is linear, we
can split the operator $\mathcal{L}$ into two parts: $\mathcal{L}_{x}$
the diffusion and drift operating in the molecule's position and orientation
coordinates, $x=(r,\theta)$, and $\mathcal{L}_{\tilde{q}}$ operating
on the atoms positions $\tilde{q}$ in the molecule's frame of reference,

\begin{equation}
\partial_{t}f(x,\tilde{q},t)=\underbrace{\mathcal{L}_{x}f(x,\tilde{q},t)}_{\substack{\text{Molecule translation }\\
\text{and rotation}
}
}+\underbrace{\mathcal{L}_{\tilde{q}}f(x,\tilde{q},t)}_{\substack{\text{Atoms diffusion in}\\
\text{molecule's frame of ref.}
}
}.\label{eq:newFP}
\end{equation}
Note it seems we increased the dimensionality of our problem by six
when adding $r$ and $\theta.$ However, we can simply get rid of
two random coordinates in vector $\tilde{q}(t)$ and recover them
through the relations in Eqs. (\ref{eq:frametransf}). This issue
will not be relevant since we will apply dimensionality reduction
techniques. 

\subsubsection{Propagators}

It is sometimes convenient to express the last equation in terms of
the propagator. To do so, we integrate Eq. \ref{eq:newFP} from $t$
to $t+\tau$,

\[
f(x,\tilde{q},t+\tau)=f(x,\tilde{q},t)+\int_{t}^{t+\tau}\mathcal{L}_{x}f(x,\tilde{q},s)ds+\int_{t}^{t+\tau}\mathcal{L}_{\tilde{q}}f(x,\tilde{q},s)ds.
\]
This defines the propagator operators $\mathcal{P}_{x}$ for the molecule's
translational and rotational diffusion and $\mathcal{P}_{\tilde{q}}$
for the diffusion of the atoms in the new coordinates, both of which
depend on the lag-time $\tau$ chosen. The equation is then

\begin{align}
f(x,\tilde{q},t+\tau) & =f(x,\tilde{q},t)+\mathcal{P}_{x}f(x,\tilde{q},t)+\mathcal{P}_{\tilde{q}}f(x,\tilde{q},t).\label{eq:prop_splitted}
\end{align}

\subsubsection{Galerkin discretization}

We call $\Omega$ the portion of the phase space spanned by the coordinates
$\tilde{q}(t)$. In this section, we use methods similar to those
presented in \citep{PrinzEtAl_JCP10_MSM1} to discretize $\Omega$
into a discrete state space. As the coordinates $\tilde{q}(t)$ are
aligned with the molecule, we will observe meta-stable regions in
the $\Omega$ phase space corresponding to the different stable structures
or conformations of the molecule. We assume there are $M$ meta-stable
regions denoted by $[s_{1},\cdots,s_{M}]$; each of these regions
defines a state or conformation of the molecule $1,\dots,M$. We then
define indicator functions on these regions/states 
\[
\chi_{i}(\tilde{q})=\begin{cases}
1 & \text{if}\quad\tilde{q}\in s_{i},\\
0 & \mathrm{else}.
\end{cases}
\]
We discretize the continuous operator $\mathcal{P}_{\tilde{q}}$ into
a discrete operator by doing a Galerkin discretization with these
indicator functions as basis functions. This is achieved by multiplying
Eq. (\ref{eq:prop_splitted}) by $\chi_{i}(x)$ and integrating over
$\Omega$

\begin{equation}
f_{i}(x,t+\tau)=f_{i}(x,t)+\int_{\Omega}\chi_{i}(\tilde{q})\mathcal{P}_{x}f(x,\tilde{q},t)d\tilde{q}+\int_{\Omega}\chi_{i}(\tilde{q})\mathcal{P}_{\tilde{q}}f(x,\tilde{q},t)d\tilde{q},\label{eq:prop_integral}
\end{equation}
where we used that
\[
f_{i}(x,t)=\int_{\Omega}\chi_{i}(\tilde{q})f(x,\tilde{q},t)d\tilde{q}
\]
is the probability of being in the meta-stable region $\chi_{i}$
at a given time. The operator $\mathcal{P}_{x}$ in the first integral
corresponds to the diffusion propagator for the translational and
rotational motion of the molecule, and we expect it to behave differently
for different conformations of the molecule. In general, $\mathcal{P}_{x}$
depends on $\tilde{q}$ through the potential (see Eq. (\ref{eq:overLoperator})),
so it cannot simply exchange order with the integration. However,
if the molecule remains on the $i^{\mathrm{th}}$meta-stable region,
we do not expected any significant change on $\mathcal{P}_{x}$, so
we can approximate it by $\mathcal{P}_{x}^{i}$, which depends on
the conformation $i$ but not on $\tilde{q}.$ Thus, we can approximate
the first integral by 

\begin{align*}
\int_{\Omega}\chi_{i}(\tilde{q})\mathcal{P}_{x}f(x,\tilde{q},t)d\tilde{q}\approx & \mathcal{P}_{x}^{i}f_{i}(x,t),
\end{align*}
We still need to further simplify the last term of Eq. \ref{eq:prop_integral}.
To do so, we focus to the eigenfunctions of $\mathcal{P}_{\tilde{q}}$,
\[
\mathcal{P}_{\tilde{q}}\psi=\lambda\psi.
\]
We then expand $f(x,\tilde{q},t)$ in these eigenfunctions, $f=\sum_{j=1}^{\infty}\langle f,\psi_{j}\rangle\psi_{j},$
and apply $\mathcal{P}_{\tilde{q}}$ 
\[
\mathcal{P}_{\tilde{q}}f=\sum_{j=0}^{\infty}\lambda_{j}\langle f,\psi_{j}\rangle\psi_{j}.
\]
As we are interested in slow processes, such as conformation switching,
we focus on larger time-scales. This means truncating this sum to
only include the $K$ slowest processes, $\sum_{j=0}^{K}\lambda_{j}\langle f,\psi_{j}\rangle\psi_{j}$,
where we assumed the indexing of the eigenvalues/eigenfunctions to
follow $\lambda_{0}\geq\lambda_{1}\geq\dots\geq\lambda_{K}$. Then
the last term of Eq. \ref{eq:prop_integral} is simply approximated
by
\[
\int_{\Gamma}\chi_{i}(\tilde{q})\mathcal{P}_{\tilde{q}}f(x,\tilde{q},t)d\tilde{q}\approx\sum_{j=1}^{K}\lambda_{j}\langle f,\psi_{j}\rangle\psi_{j}^{i}V_{i}
\]
where $\psi_{j}^{i}=\frac{1}{V_{i}}\int_{\Omega}\chi_{i}(\tilde{q})\psi_{j}d\tilde{q}$
is the average value of the $j^{\mathrm{th}}$ eigenfunction over
the $i^{\mathrm{th}}$ state, and $V_{i}$ is the volume of the $i^{\mathrm{th}}$
state in the phase space $\Omega$. The inner product integral can
be then approximated by piecewise constant values along the states,
$\langle f,\psi_{j}\rangle=\int_{\text{\ensuremath{\Omega}}}f(x,\tilde{q},t)\psi_{j}(\tilde{q})d\tilde{q}\approx\sum_{k=1}^{M}f_{k}(x,t)\psi_{j}^{k}$,
so the last term is simply 
\[
\int_{\Gamma}\chi_{i}(\tilde{q})\mathcal{P}_{\tilde{q}}f(x,\tilde{q},t)d\tilde{q}\approx\sum_{k=1}^{M}f_{k}(x,t)\underbrace{V_{i}\sum_{j=1}^{K}\lambda_{j}\psi_{j}^{k}\psi_{j}^{i}}_{\alpha_{ik}},
\]
Consequently, Eq. \ref{eq:prop_integral} is simplified to

\[
f_{i}(x,t+\tau)=f_{i}(x,t)+\mathcal{P}_{x}^{i}f_{i}(x,t)+\sum_{k=1}^{M}\alpha_{ik}f_{k}(x,t).
\]
We refer to this as the discrete-time hybrid switching diffusion model,
which can also be written in its vector form
\begin{equation}
\bar{f}(x,t+\tau)=\bar{f}(x,t)+\mathcal{P}_{x}\bar{f}(x,t)+\mathbb{P}\bar{f}(x,t),\label{eq:prop_final}
\end{equation}
where $\bar{f}(x,t)=[f_{1}(x,t),\cdots,f_{M}(x,t)]$ is a vector with
functions as entries, and it corresponds to a partially discrete approximation
to the probability density $f$ from Eq. \ref{eq:newFP}. Each function
in the vector $f_{i}(x,t)$ corresponds to the probability density
of observing the molecule in conformation $i$ at position and orientation
$x$ at time $t$. The term $\mathcal{P}_{x}\bar{f}(x,t)=[\mathcal{P}_{x}^{1}f_{1}(x,t),\cdots,\mathcal{P}_{x}^{M}f_{M}(x,t)]$
is the vector of diffusion propagators, each corresponding to a different
conformation. The last term is a discrete-time MSM, where $\mathbb{P}$
is an $M\times M$ transition probability matrix with entries $\alpha_{ik}$
representing the transition probabilities between the states or conformations.
As probability must be conserved, the columns of the transition matrix
$\mathbb{P}$ must sum to one, $\sum_{i=1}^{M}\alpha_{ik}=1$. This
can be achieved by renormalizing the rates $\alpha_{ik}$ by multiplying
them by a renormalization factor $w_{i}$, such that the condition
that needs to be satisfied is $\sum_{i=1}^{M}w_{i}\alpha_{ik}=1$.
This yields a system of $M$ unknowns with $M$ equations, so we can
solve for every $w_{i}$. This renormalization is valid because it
corresponds to multiplying each eigenfunction by a constant factor,
which remains an eigenfunction. 

In practice, we extract the rates in $\mathbb{P}$ from data using
the methodology from \citep{PrinzEtAl_JCP10_MSM1} and the core MSMs
approach from \citep{SchuetteEtAl_JCP11_Milestoning}. We should also
point out it is possible to derive conservative discretizations by
smartly choosing non-constant basis functions $\chi(\tilde{q})$,
such as in the discontinuous Galerkin methods.

\subsubsection{The hybrid switching diffusion model}

The model we derived in Eq. \ref{eq:prop_final} can be framed in
a continuous time context. If we divide the equation \ref{eq:prop_final}
by $\tau$ and take the limit $\tau\rightarrow0$, we obtain an equation
similar to the original Fokker-Planck equation (Eq. \ref{eq:newFP}),

\begin{equation}
\partial_{t}\bar{f}(x,t)=\underbrace{\mathcal{D}\bar{f}(x,t)}_{\substack{\text{Molecule translation}\\
\text{and rotation}
}
}+\underbrace{\mathbb{Q}\bar{f}(x,t)}_{\substack{\text{Continuous-time}\\
\text{MSM}
}
},\label{eq:coupleddiff}
\end{equation}
where $\mathbb{Q}$ is now a transition-rate matrix modeling the molecule's
conformation changes. The molecule's translational and rotational
diffusion is modeled by $\mathcal{D}\bar{f}=[\mathcal{D}_{1}f_{1},\cdots,\mathcal{D}_{M}f_{M}]$.
These entries are obtained as the limit, $\mathcal{D}_{i}f_{i}=\lim_{\tau\rightarrow0}\mathcal{P}_{x}f_{i}/\tau$.
We refer to this model simply as the hybrid switching diffusion model.
The main differences with respect to Eq. \ref{eq:newFP} are that
here the phase space spanned by $q$ was discretized into meta-stable
regions and that the dynamics of fast processes have been filtered
out. 

Alternatively one could also write the following SDE to describe the
individual stochastic trajectories, whose ensemble distribution dynamics
are governed by Eq. \ref{eq:diffMS_01}

\begin{equation}
dX(t)=\mu(X(t),\eta(t),t)dt+\sigma(X(t),\eta(t),t)dW(t),\label{eq:diffMS_SDE}
\end{equation}

where $X(t)$ corresponds to the postion and orientation of the molecule,
the drift $\mu$ and the diffusion $\sigma$ depend on the form of
the diffusion operator $\mathcal{D}$, as well as on the current conformation
of the molecule $\eta(t),$with $\eta(t)$ only accepting discrete
values. The random conformation at a given time for one trajectory
can be computed using a Gillespie-type algorithm based on the rates
in matrix $\mathbb{Q}$. 

Equation \ref{eq:coupleddiff} is an example of a hybrid switching
diffusion process. This is a well-defined stochastic process, and
it has been thoroughly studied in \citep{mao2006stochastic,yin2010hybrid}.
These hybrid switching diffusions \citep{yin2010hybrid} models are
also known in the mathematics community as diffusion processes with
Markovian switching \citep{mao2006stochastic} or coupled diffusion
models. They are called 'hybrid' due to the coexistence of continuous
dynamics and discrete events in the same process.

The switching diffusion model can be formulated using discrete time
(Eq. \ref{eq:prop_final}) or continuous time (Eq. \ref{eq:coupleddiff}),
depending of which formulation is more convenient for the corresponding
application. The translational and rotational diffusion of the molecule
is modeled by either $\mathcal{P}_{x}\bar{f}$ or $\mathcal{D}\bar{f}$,
while the switching between different states or conformations of the
molecule is modeled by a discrete-time or continuous-time MSM, $\mathbb{P}$
and $\mathbb{Q}$, respectively. 

\subsection{Hybrid switching diffusion for one molecule \label{app:onemolCD}}

In a more realistic setting, we need to provide a more robust model
for the dynamics. Consider the dynamics of one molecule $A$, with
$N$ atoms. Assume molecule $A$ has $M$ conformations. The position
and momentum of the $N$ atoms is given by $q(t)$ and $p(t)$, respectively.
Analogously to the previous example, we assume the dynamics of the
stochastic dynamical system are governed by the Langevin equation,
\begin{align}
dq(t)= & p(t)dt,\nonumber \\
dp(t)= & -\nabla_{q}U(q(t))dt-\Gamma p(t)dt+\Sigma dW(t),\label{eq:oneMLangevin}
\end{align}
where to simplify notation, we assumed the mass of all atoms to be
one; $U(q(t))$ is the interaction potential; $\Gamma$ is a diagonal
matrix with the damping coefficients $\gamma_{i}$ along the diagonal;
$\Sigma$ is a matrix that correlates the noise vector, and $W(t)$
is a vector of standard Brownian motions. The corresponding Fokker-Planck
equation for the probabilistic dynamics in phase space is then given
by 
\begin{align}
\partial_{t}f(q,p,t) & =\mathcal{L}_{A}f(q,p,t)\label{eq:generatorLang}
\end{align}
where $f(q(t),p(t),t)$ is the probability density function of the
system as a function of ($q(t),p(t),t$); the Langevin operator is
$L_{A}=\left[\nabla_{p}\cdot D\nabla-p\cdot\nabla_{q}+\nabla_{q}U(q)\cdot\nabla_{p}+\gamma p\cdot\nabla_{p}+\mathrm{tr}(\Gamma)\right]$,
$D=\frac{1}{2}\Sigma\Sigma^{\mathrm{T}}$, $\mathrm{tr}(\Gamma)$
is the trace of $\Gamma$. The operators $\nabla_{q}$, $\nabla_{p}$
are the gradient operators with respect to the position and momentum
coordinates, respectively. Note if the noise is uncorrelated, $\Sigma$
and $D$ are diagonal matrices with $\sigma_{ii}=\sqrt{2k_{B}T\gamma_{i}}$
and $D_{ii}=\sigma_{ii}^{2}/2$, along the diagonal.

Following Section \ref{app:CD1stderivation} and Eq. \ref{eq:frametransf},
we change again the frame of reference to the center of mass of $A$,
$r_{A}=\frac{1}{N}\sum_{i=1}^{N}q_{i}(t)$, and we rotate the frame
by $\theta_{A}^{-1}$ to align it to the molecule. Similarly, we need
to transform the momentum coordinates, 

\[
\tilde{p}_{i}(t)=R_{\theta}^{-1}p_{i}(t),\qquad p_{A}(t)=\frac{1}{N}\sum_{i=1}^{N}p_{i}(t),
\]
where $p_{A}$ correspond to the velocity of the center of mass. The
molecule's angular momentum in the center of mass reference is given
by $\omega_{A}$, which is a function $p$. Note the translation to
the center of mass is invariant for the velocities. We can again derive
SDEs for $r_{A}$, $p_{A}$, $\theta_{A}$ and $\omega_{A}$ and for
all the atoms variables in the new coordinates $\tilde{q}_{i}(t)$
and $\tilde{p}_{i}(t)$. The Fokker-Planck equation can be now conveniently
separated into an operator acting on the molecule's position, velocity,
orientation and angular momentum $z=(r_{A},p_{A},\theta_{A},\omega_{A})$,
and another operator acting on the individual atoms position and momentum,$q$
and $p$, in the molecule's reference system,

\begin{equation}
\partial_{t}f(z,\tilde{q},\tilde{p},t)=\underbrace{\mathcal{L}_{D}f(z,\tilde{q},\tilde{p},t)}_{\substack{\text{Molecule translation }\\
\text{and rotation}
}
}+\underbrace{\mathcal{L}_{qp}f(z,\tilde{q},\tilde{p},t)}_{\substack{\text{Atoms diffusion in}\\
\text{molecule's frame of ref.}
}
}.\label{eq:newFP00-2}
\end{equation}
Following the same methodology as in Appendix \ref{app:CD1stderivation},
we obtain the hybrid switching diffusion model for one molecule,

\[
\partial_{t}\bar{f}(z,t)=\underbrace{D_{A}\bar{f}(z,t)}_{\substack{\text{Molecule translation}\\
\text{and rotation}
}
}+\underbrace{Q_{A}\bar{f}(z,t)}_{\mathrm{\substack{\text{Continuous-time}\\
\text{MSM}
}
}}.
\]
where $\bar{f}(z,t)=[f_{1}(z,t),\cdots,f_{M}(z,t)]$. In the cases
where the time-scales we are interested in are larger than the autocorrelation
decay times of $p_{A}$and $\omega_{A}$, the first operator can be
approximated by overdamped operators for the diffusion and rotation,
$D_{A}\approx\mathcal{D}_{A}$, the momentum coordinates become irrelevant,
and the transition rate matrix $Q_{A}$ is projected into the corresponding
space resulting in $\mathbb{Q}_{A}$. This simplifies the system to

\begin{equation}
\partial_{t}\bar{f}(x,t)=\mathcal{D}_{A}\bar{f}(x,t)+\mathbb{Q}_{A}\bar{f}(x,t),\label{eq:oneMolCD}
\end{equation}

where $x=(r_{A},\theta_{A})$. This equation is exactly of the same
form as Eq. \ref{eq:coupleddiff}. Note if the time-scales of interest
are even larger than the time-scales for conformational changes described
by $Q_{A}$, the MSM becomes irrelevant, the vector $\bar{f}(x,t)$
has only one component, and the diffusion properties of the different
conformations are averaged out by $\mathcal{D}_{A}$, recovering the
Fokker-Planck equation for one diffusing and rotating particle. Although
not yet explored, more complex derivations formally taking into account
the solvent could lead to interesting models with memory kernels \citep{ford1965statistical,zwanzig1973nonlinear}.

\subsection{Hybrid switching diffusion for two interacting molecules \label{app:twomolCD}}

Consider the dynamics of two molecules $A$ and $B$, each one with
$N_{A}$and $N_{B}$ atoms, respectively. Assume molecule $A$ has
$M_{A}$ conformations and $B$ has $M_{B}$ conformations, and they
can bound in $M_{C}$ different meta-stable configurations. The position
and momentum of every atom is given by $q(t)=[q_{A}(t),q_{B}(t)]$
and $p(t)=[p_{A}(t),p_{B}(t)]$ ($q_{X}(t)$ and $p_{X}(t)$ denote
the positions and momentum of all the atoms in molecule $X$). Analogous
to Appendix \ref{app:onemolCD}, we assume all the masses are one
and that the dynamics of the stochastic dynamical system are governed
by the Langevin equation
\begin{align*}
dq(t) & =p(t)dt,\\
dp(t) & =-\nabla_{q}U(q(t))dt-\Gamma p(t)dt+\Sigma dW(t)
\end{align*}
The corresponding Fokker-Planck equation is given by, 
\begin{align}
\partial_{t}f(q,p,t) & =L_{AB}f(q,p,t)\label{eq:generatorLang2}\\
L_{AB} & =\frac{1}{2}\nabla_{p}\cdot D\nabla_{p}-p\cdot\nabla_{q}+\nabla_{q}U(q)\cdot\nabla_{p}\nonumber \\
 & +\gamma p\cdot\nabla_{p}+\mathrm{tr}(\Gamma)\nonumber 
\end{align}
where $f(q(t),p(t),t)$ is the probability density of the system as
a function of ($q(t),p(t),t$), $D=\frac{1}{2}\Sigma\Sigma^{\mathrm{T}}$
, and $L_{AB}$ is the infinitesimal generator of the process. The
potential function can be rewritten as a sum of the independent potentials
corresponding to each molecule plus an interaction term 
\[
U(q)=U_{A}(q_{A})+U_{B}(q_{B})+\Phi(r_{AB})U_{AB}(q_{A},q_{B}),
\]
where $r_{AB}=|r_{B}-r_{A}|$ is the distance between the two molecules,
with $r_{A}$ and $r_{B}$ the centers of mass of $A$ and $B$. The
function $\Phi(r_{AB})$ varies between $0$ and $1$ to weight how
strong is the interaction as a function of $r_{AB}$, so when the
molecules are close ($r_{AB}\ll$1), $\Phi(r_{AB})=1$ and when they
are far apart ($r_{AB}\gg$1), $\Phi(r_{AB})=0$. This divides our
phase space in three regimes: the non-interacting regime ($\Phi(r_{AB})=0$),
the transition regime ($0<\Phi(r_{AB})<1$), and the bound regime
($\Phi(r_{AB})=1$). 

\subsubsection{\textcolor{black}{Non-interacting regime}}

In this regime $\Phi(r_{AB})=0$, so we only have the first two terms
in the potential. Furthermore, when $r_{AB}$ is large, the two molecules
dynamics are independent, so the joint probability can be separated
as the product of the probability densities corresponding to each
molecule $f_{AB}(q,p,t)=f_{A}(q_{A},p_{A},t)f_{B}(q_{B},p_{B},t)$.
Taking this into consideration we can rewrite each term of Eq. \ref{eq:generatorLang2}
as a term acting on the coordinates ($q_{A},p_{A}$) or ($q_{B},p_{B}$),
\begin{gather*}
\partial_{t}(f_{A}f_{B})=\left[\frac{1}{2}\nabla_{p_{A}}\cdot D_{A}\nabla_{p_{A}}-p_{A}\cdot\nabla_{q_{A}}+\nabla_{q_{A}}V_{A}(q_{A})\cdot\nabla_{p_{A}}\right.\\
+\gamma p_{A}\cdot\nabla_{p_{A}}+\mathrm{tr}(\Gamma_{A})+\frac{1}{2}\nabla_{p_{B}}\cdot D_{B}\nabla_{p_{B}}-p_{B}\cdot\nabla_{q_{B}}\\
+\nabla_{q_{B}}V_{B}(q_{B})\cdot\nabla_{p_{B}}+\gamma p_{B}\cdot\nabla_{p_{B}}+\mathrm{tr}(\Gamma_{B})\Bigr]f_{A}f_{B},
\end{gather*}
where $D_{X}$ is a block matrix of $D$ corresponding to the molecule
$X$, $\Gamma_{x}$ is the block matrix of $\Gamma$ corresponding
to $X$, and we omitted variable dependencies to ease notation. Notice
that $\mathrm{tr}(\Gamma)=\mathrm{tr}(\Gamma_{A})+\mathrm{tr}(\Gamma_{B})$.
The left hand side can be simply expanded as $f_{A}\partial_{t}f_{B}+f_{B}\partial_{t}f_{A}$,
where $\partial t$ denotes partial derivative with respect to $t$,
and we can divide the whole equation by $f_{A}f_{B}$ to obtain, 
\begin{gather*}
\frac{\partial_{t}f_{A}}{f_{A}}+\frac{\partial_{t}f_{B}}{f_{B}}=\\
\frac{1}{f_{A}}\left[\frac{1}{2}\nabla_{p_{A}}\cdot D_{A}\nabla_{p_{A}}-p_{A}\cdot\nabla_{q_{A}}+\nabla_{q_{A}}V_{A}(q_{A})\cdot\nabla_{p_{A}}\right.\\
+\gamma p_{A}\cdot\nabla_{p_{A}}+\mathrm{tr}(\Gamma_{A})\Bigr]f_{A}+\\
\frac{1}{f_{B}}\left[\frac{1}{2}\nabla_{p_{B}}\cdot D_{B}\nabla_{p_{B}}-p_{B}\cdot\nabla_{q_{B}}+\nabla_{q_{B}}V_{B}(q_{B})\cdot\nabla_{p_{B}}\right.\\
+\gamma p_{B}\cdot\nabla_{p_{B}}+\mathrm{tr}(\Gamma_{B})\Bigr]f_{B}.
\end{gather*}
Note we have only functions of $(q_{A},p_{A})$ or $(q_{B},p_{B})$
and constants. We can move the terms such that the left hand side
is only a function of $(q_{A},p_{A})$ and the right hand side just
a function of $(q_{B},p_{B})$, such that for the equality to hold
they must be equal to a constant. Considering these are equations
that described probabilities, the only reasonable choice for this
constant is zero, so
\begin{align*}
\partial_{t}f_{A} & =\left[\frac{1}{2}\nabla_{p_{A}}\cdot D_{A}\nabla_{p_{A}}-p_{A}\cdot\nabla_{q_{A}}+\nabla_{q_{A}}V_{A}(q_{A})\cdot\nabla_{p_{A}}\right.\\
 & +\gamma p_{A}\cdot\nabla_{p_{A}}+\mathrm{tr}(\Gamma_{A})\Bigr]f_{A}\\
\partial_{t}f_{B} & =\left[\frac{1}{2}\nabla_{p_{B}}\cdot D_{B}\nabla_{p_{B}}-p_{B}\cdot\nabla_{q_{B}}+\nabla_{q_{B}}V_{B}(q_{B})\cdot\nabla_{p_{B}}\right.\\
 & +\gamma p_{B}\cdot\nabla_{p_{B}}+\mathrm{tr}(\Gamma_{B})\Bigr]f_{B},
\end{align*}
which is by definition the infinitesimal generator for the independent
dynamics of each of the two individual molecules, 
\begin{align}
\partial_{t}f_{A}(q_{A},p_{A},t) & =\mathcal{L}_{A}f_{A}(q_{A},p_{A},t),\nonumber \\
\partial_{t}f_{B}(q_{B},p_{B},t) & =\mathcal{L}_{B}f_{B}(q_{B},p_{B},t).\label{eq:subGeneratorsLang}
\end{align}
This further means that we have two independent Langevin equations
for each molecule. Following the methodology from Appendix \ref{app:onemolCD}
and assuming the diffusion dynamics are accurately approximated by
overdamped Langevin dynamics, see Eq. \ref{eq:oneMolCD}, we can then
write a hybrid switching diffusion model for each of the molecules

\begin{align}
\partial_{t}\bar{f_{A}}(x_{A},t) & =\mathcal{D}_{A}\bar{f_{A}}(x_{A},t)+\mathbb{Q}_{A}\bar{f_{A}}(x_{A},t),\nonumber \\
\partial_{t}\bar{f_{B}}(x_{B},t) & =\mathcal{D}_{B}\bar{f_{B}}(x_{B},t)+\mathbb{Q}_{B}\bar{f_{B}}(x_{B},t),\label{eq:2molCD_far}
\end{align}

where $x_{i}=(r_{i},\theta_{i})$ is the position and orientation
of molecule $i$. The vectors $\bar{f_{A}}$ and $\bar{f_{B}}$ have
the dimensions of the individual conformations $M_{A}$ and $M_{B}$,
respectively. The transition rate matrices $\mathbb{Q}_{A}$ and $\mathbb{Q}_{B}$
have dimensions $M_{A}\times M_{A}$ and $M_{B}\times M_{B}$, respectively.
Although at this point we can describe each molecule individually,
it will be useful to show the state of the full system is simply given
by the tensor product $\bar{f}_{AB}=\bar{f_{A}}\otimes\bar{f_{B}}$.
Its solution is the tensor product of the individual solutions,
\[
\bar{f}_{AB}(t)=e^{t\mathcal{D}_{A}}\otimes e^{t\mathcal{D}_{B}}\otimes e^{t(\mathbb{Q}_{A}\oplus\mathbb{Q}_{B})}\bar{f}_{0},
\]
with $\bar{f}_{0}=\bar{f}_{A}(0)\otimes\bar{f}_{B}(0)$. This way
we can define the coupled diffusion coefficient as $e^{tD_{AB}}=e^{t\mathcal{D}_{A}}\otimes e^{t\mathcal{D}_{B}}$.
This last equation describes the diffusion of the system in terms
of the diffusion of the centers of mass and orientation of the two
molecules, and it describes its state in terms of the continuous-time
MSM given by the transition rate matrix

\[
\mathbb{Q}_{AB}=\mathbb{Q}_{A}\oplus\mathbb{Q}_{B}.
\]
This type of sum is called the Kronecker sum. Note that in the case
of the discontinuous-time hybrid switching diffusion model, the resulting
MSM will simply be given by the tensor product of the transition matrices
$\mathbb{P}_{AB}=\mathbb{P}_{A}\otimes\mathbb{P}_{B}$, since $\bar{f}_{AB}(t)=e^{t\mathcal{D}_{A}}\otimes e^{t\mathcal{D}_{B}}\otimes\left(\mathbb{P}_{A}\otimes\mathbb{P}_{B}\right)\bar{f}_{0}.$

\subsubsection{Bound regime}

In this case, as $\Phi(r_{AB})=1$, we are not able to uncouple Eq.
\ref{eq:generatorLang2} into the independent molecules dynamics,
so we have no alternative but to describe the dynamics of the joint
complex. Therefore we apply the same change of variables as in Appendix
\ref{app:onemolCD} for the position and momentum of the joint complex,
$C$. Following the same methodology as in Appendix \ref{app:onemolCD}
and assuming the diffusion dynamics can be approximated by overdamped
Langevin dynamics, we obtain the hybrid switching diffusion model
for the two coupled molecules in the small separation regime,

\begin{align}
\partial_{t}\bar{f_{C}}(x,t) & ==\underbrace{\mathcal{D}_{C}\bar{f_{C}}(x,t)}_{\substack{\text{\ensuremath{C} translation }\\
\text{and rotation}
}
}+\underbrace{\mathbb{Q}_{C}\bar{f_{C}}(x,t)}_{\substack{\text{MSM for }\\
\text{\ensuremath{C} complex}
}
},\label{eq:2molCD_close}
\end{align}

with $z=(r_{C},\theta_{C})$, corresponding to the $C$-complex center
of mass position and orientation; and where $\bar{f_{C}}$ has the
dimensions of all the possible bound configurations between the two
molecules $M_{C}$, and $\mathbb{Q}_{c}$ has dimensions $M_{C}\times M_{C}$. 

\subsubsection{Transition regime}

We would like to have a regime that allows us to transition between
the non-interacting and bound regime from Eqs. \ref{eq:2molCD_far}
and \ref{eq:2molCD_close}, respectively. The interactions are encoded
in the transition rate matrix $\mathbb{Q}$ and should depend on the
separation between the two molecules, $r_{AB}=|r_{B}-r_{A}|$, as
well as in their relative orientation, $\theta_{AB}=\theta_{B}\theta_{A}^{-1}$.
Therefore, the transition rate matrix $\mathbb{Q}(x_{AB})$ depends
on $x_{AB}=(r_{AB},\theta_{AB})$. In the non-interacting regime,
Eq. \ref{eq:2molCD_far} corresponds to $r_{AB}\gg1$such that the
dependence on $x_{AB}$ is zero and $\mathbb{Q}(x_{AB})\rightarrow\mathbb{Q}_{A}\oplus\mathbb{Q}_{B}$.
In the bound regime, the dependence on $x_{AB}$ is again lost, and
all transitions are governed by $\mathbb{Q}_{C}$, unless there is
a dissociation event. We thus need a transition regime, where the
rates depend on $x_{AB}$ and the system can transition from a dissociated
state to a bound state and vice versa. This can be achieved by incorporating
the two limiting models we just obtained into a larger switching diffusion
process that can interact through the rate matrices $\mathbb{Q}_{AB\rightarrow C}$
and $\mathbb{Q}_{C\rightarrow AB}$. This can be done by writing the
transition matrix $\mathbb{Q}$as a block matrix

\begin{align*}
\frac{\partial\bar{f}(x)}{\partial t} & =\mathcal{D}\bar{f}(x)+\mathbb{Q}(x_{AB})\bar{f}(x),\\
\mathbb{Q}(x_{AB}) & =\left(\begin{array}{c|c}
\mathbb{Q}_{AB} & \mathbb{Q}_{C\rightarrow AB}\\
\hline \mathbb{Q}_{AB\rightarrow C} & \mathbb{Q}_{C}
\end{array}\right),
\end{align*}

where $\bar{f}=(\bar{f}_{AB},\bar{f}_{C})$, with $\bar{f}_{AB}=\bar{f_{A}}\otimes\bar{f_{B}}$
the vector of all the unbound states given by the tensor product of
independent states of $A$ and $B$, and $\bar{f}_{C}$ the vector
of all the bound states of the $C-$complex. The diffusion term $\mathcal{D}\bar{f}(x)$
can in principle also depend on $x_{AB}$; however, it can be approximated
by the combined diffusions of the two limiting cases $\mathcal{D}\bar{f}=[\mathcal{D}_{AB}\bar{f}_{AB},\mathcal{D}_{C}\bar{f}_{C}]$
since the interactions are already encoded in $\mathbb{Q}$. Overall,
$\bar{f}$ has dimension $M=M_{A}M_{B}+M_{C}$ and $\mathbb{Q}$ has
dimensions $M\times M$. In the non-interacting regime $\mathbb{Q}_{AB}=\mathbb{Q}_{A}\oplus\mathbb{Q}_{B}$
and $\mathbb{Q}_{AB\rightarrow C}=0$. In the transition regime, $\mathbb{Q}_{AB\rightarrow C}\neq0$
and $\mathbb{Q}_{AB}$ has to change as well, both dependent on $x_{AB}$.
The transition rate functions $\mathbb{Q}_{AB}$ together with the
transition rate functions $\mathbb{Q}_{AB\rightarrow C}$ should always
have their combined columns in $\mathbb{Q}$ sum to zero, for any
given $x_{AB}$, ensuring that $\mathbb{Q}$ remains a transition
rate matrix. In the bound regime, the particles can transition back
to the transition regime through $\mathbb{Q}_{C\rightarrow AB}$,
which together with $\mathbb{Q}_{C}$ should also have their combined
columns sum to zero. This implies that the matrix $\mathbb{Q}_{C}$
is a renormalized version of the one in Eq. \ref{eq:2molCD_close}.

From a practical point of view, it is unlikely that one could accurately
extract these rate functions in the transition region from MD data.
However, one can be less ambitious and try to extract a piecewise
constant approximation of these rate functions. This requires deriving
a smart discretization of the six dimensional space spanned by $x_{AB}$;
one possible discretization is the one presented in the methods section.

\begin{figure*}
\centering 
\textbf{a.}\includegraphics[width=0.33\textwidth]{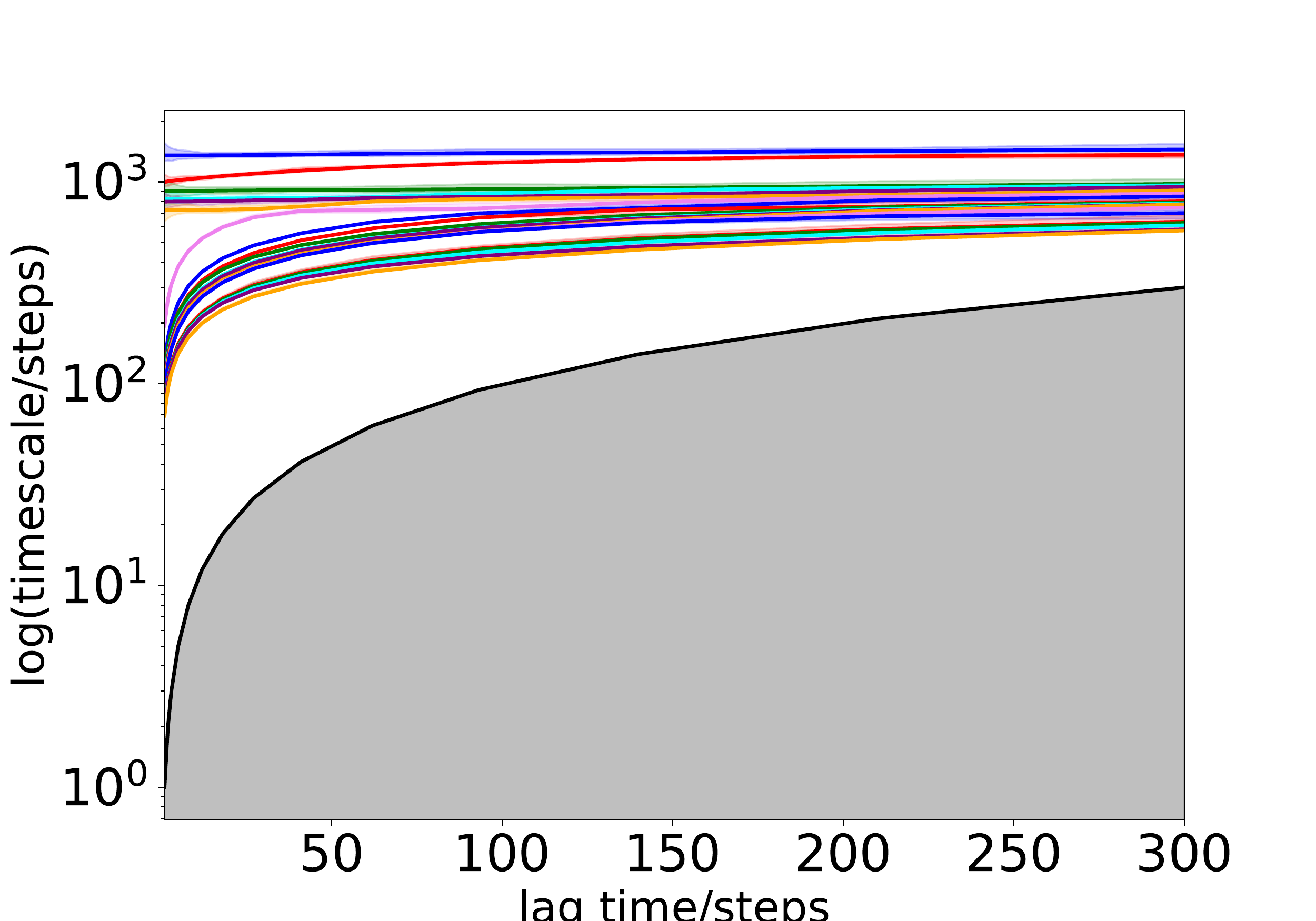}\textbf{b.}\includegraphics[width=0.33\textwidth]{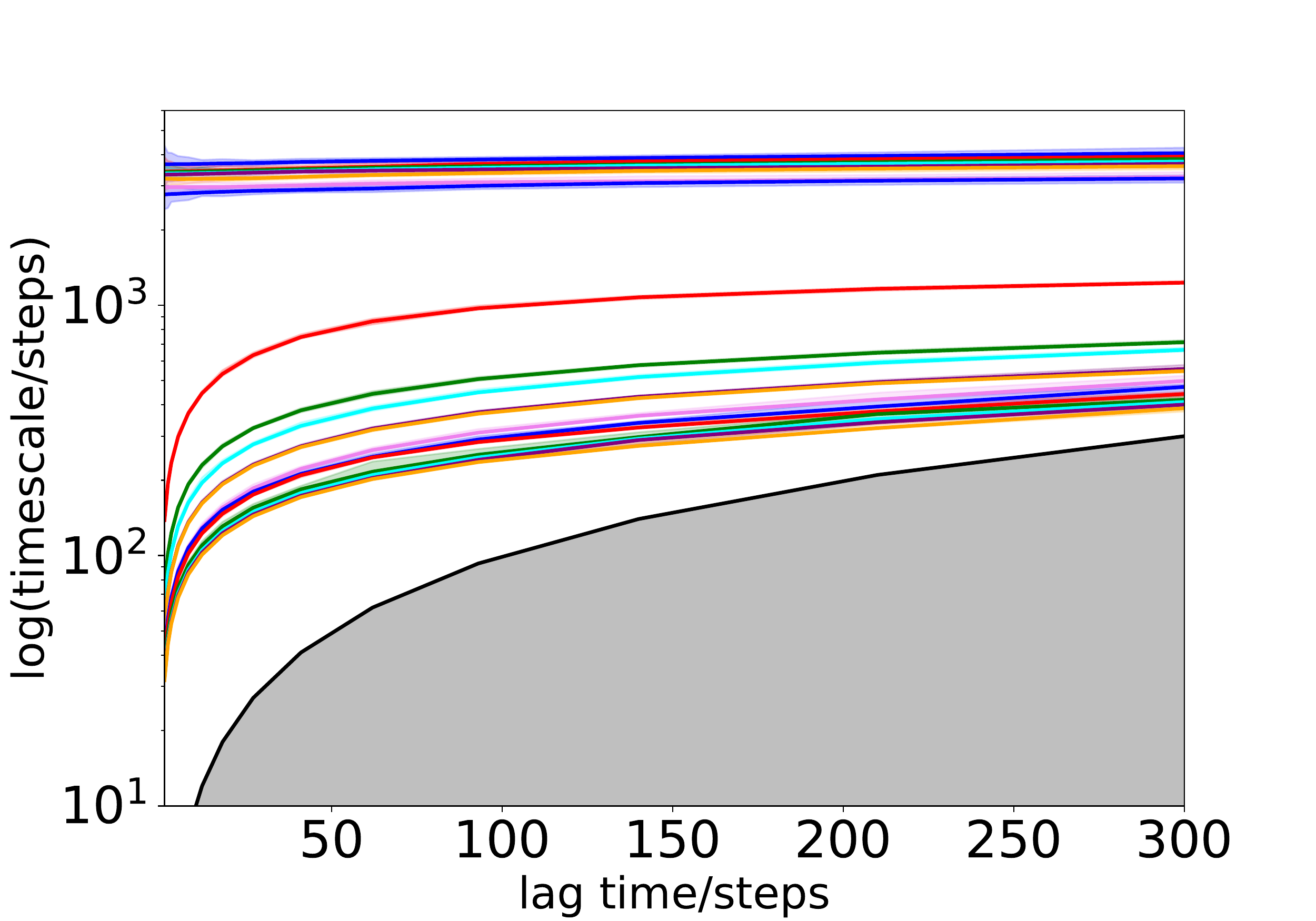}\textbf{c.\includegraphics[width=0.33\textwidth]{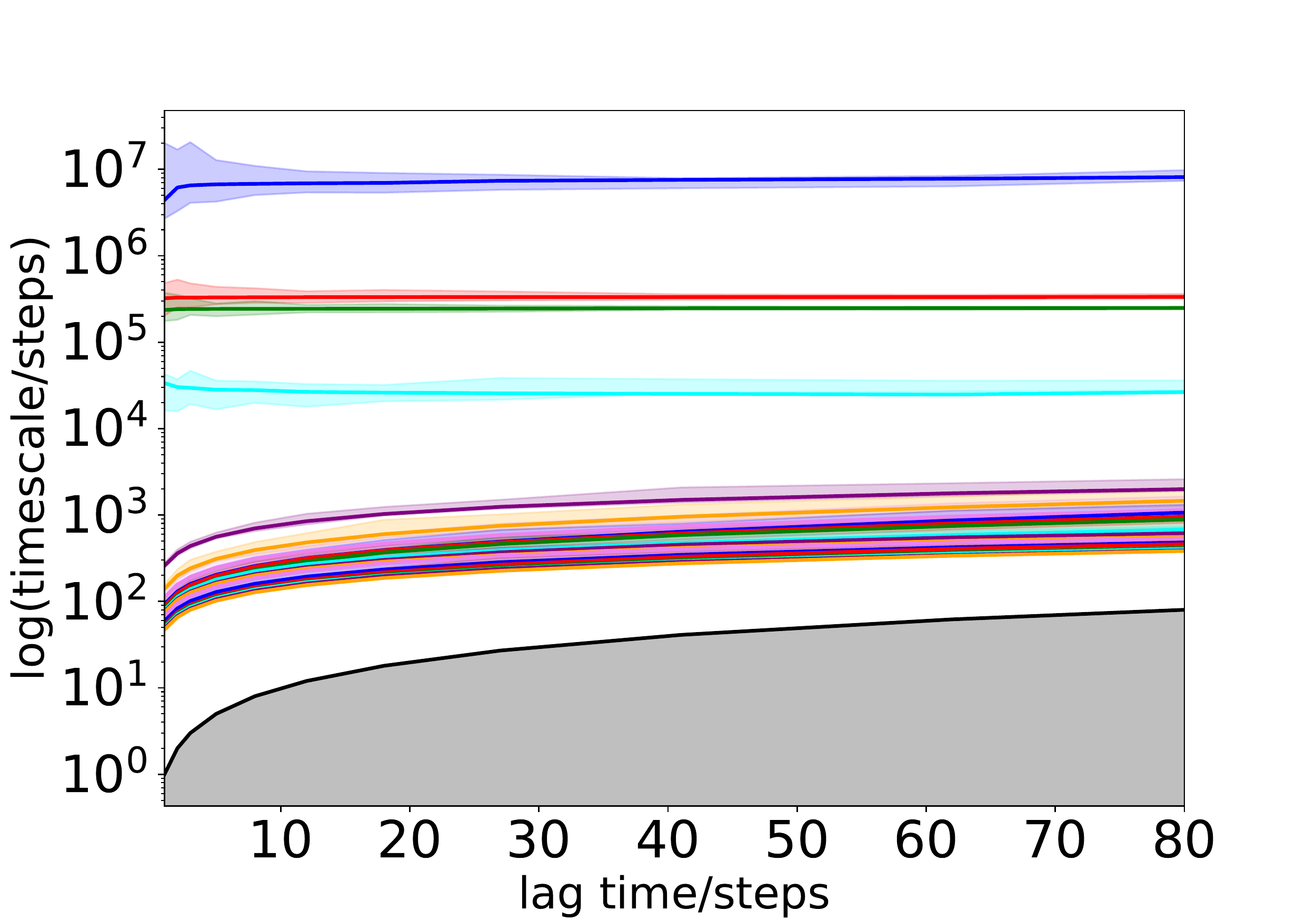}}
\caption{Implied time-scales of MSMs including all the bound states and transition
states. \textbf{a.} Implied time-scales of the protein-protein system
from Section \ref{subsec:protein-proteinMSMRD}. \textbf{b.} Implied
time-scales of the dimer system from Section \ref{subsec:implementation1}.
\textbf{c.} Implied time-scales of the dimer system used for the multiparticle
MSM/RD from Section \ref{sec:multiparticle}.}

\label{fig:ITSboth}
\end{figure*}

\begin{algorithm*}
\texttt{\textbf{\footnotesize{}Input:}}\texttt{\footnotesize{} position
and orientation of molecule A and molecule B, maximum time $T$, time
step $\delta t$.}{\footnotesize\par}
\begin{itemize}
\item \texttt{\footnotesize{}While $t<T$:}{\footnotesize\par}
\begin{enumerate}
\item \texttt{\footnotesize{}Calculate relative position $r_{AB}$ and relative
orientation $\theta_{AB}$; and get current regime.}{\footnotesize\par}
\item \texttt{\footnotesize{}If in non-interacting regime:}{\footnotesize\par}
\begin{enumerate}
\item \texttt{\footnotesize{}Propagate MSM by $\delta t$ using the transition
rate matrix $\mathbb{Q}_{AB}=\mathbb{Q}_{A}\oplus\mathbb{Q}_{B}$,
or $\mathbb{Q}_{A}$ and $\mathbb{Q}_{B}$ independently.}{\footnotesize\par}
\item \texttt{\footnotesize{}Propagate diffusion by $\delta t$ following
Eq. \ref{eq:overdGeneral} with $k=A,B$.}{\footnotesize\par}
\item \texttt{\footnotesize{}If $\left\Vert r_{AB}\right\Vert <R$: switch
to transition regime}{\footnotesize\par}
\end{enumerate}
\item \texttt{\footnotesize{}Else if in transition regime:}{\footnotesize\par}
\begin{enumerate}
\item \texttt{\footnotesize{}Propagate MSM by $\delta t$ using the piecewise
constant approximation of the transition rate matrices $\mathbb{Q}_{AB}$
and $\mathbb{Q}_{AB\rightarrow C}$.}{\footnotesize\par}
\item \texttt{\footnotesize{}If transition towards a bound state $C$ happened:}{\footnotesize\par}
\begin{enumerate}
\item \texttt{\footnotesize{}Switch to bound regime}{\footnotesize\par}
\item \texttt{\footnotesize{}Set position and orientation of $C$ compound
as average between $A$ and $B$.}{\footnotesize\par}
\end{enumerate}
\item \texttt{\footnotesize{}Else:}{\footnotesize\par}
\begin{enumerate}
\item \texttt{\footnotesize{}Propagate diffusion by $\delta t$ following
Eq. \ref{eq:overdGeneral} with $k=A,B$.}{\footnotesize\par}
\item {\footnotesize{}If }\texttt{\footnotesize{}$\left\Vert r_{AB}\right\Vert \geq R$: switch
to non-interacting regime}{\footnotesize\par}
\end{enumerate}
\end{enumerate}
\item \texttt{\footnotesize{}Else if in bound regime:}{\footnotesize\par}
\begin{enumerate}
\item \texttt{\footnotesize{}Propagate MSM by $\delta t$ using the transition
rate matrices $\mathbb{Q}_{C}$ and $\mathbb{Q}_{C\rightarrow AB}$.}{\footnotesize\par}
\item \texttt{\footnotesize{}If transition towards an unbound transition
state happened:}{\footnotesize\par}
\begin{enumerate}
\item \texttt{\footnotesize{}Switch to transition regime}{\footnotesize\par}
\item \texttt{\footnotesize{}Set position and orientation of $A$ and $B$
corresponding to the unbound transition state.}{\footnotesize\par}
\end{enumerate}
\item \texttt{\footnotesize{}Else:}{\footnotesize\par}
\begin{enumerate}
\item \texttt{\footnotesize{}Propagate diffusion by $\delta t$ following
Eq. \ref{eq:overdBound}.}{\footnotesize\par}
\end{enumerate}
\end{enumerate}
\item {\footnotesize{}Set $t\leftarrow t+\delta t$}{\footnotesize\par}
\end{enumerate}
\end{itemize}
\caption{Main algorithm for the MSM/RD scheme. Note the MSMs used in the algorithm
use a constant $\delta t$ to avoid synchronization issues. This is
particularly convenient since we will use discrete-time MSM in our
implementations with lag-times that are multiples of the simulation
time step. It is possible to modify the algorithm to use continuous-time
MSMs and propagate them with a Gillespie algorithm. \label{alg:MSM/RDmain}}
\end{algorithm*}

\begin{algorithm*}
\texttt{\textbf{\footnotesize{}Input:}}\texttt{\footnotesize{} position
and orientation of all molecules, maximum time $T$, time step $\delta t$.}{\footnotesize\par}
\begin{itemize}
\item \texttt{\footnotesize{}While $t<T$:}{\footnotesize\par}

\texttt{\footnotesize{}For all possible pairs of particles (we label
a pair as particle $A$ and particle $B$):}{\footnotesize\par}
\begin{enumerate}
\item \texttt{\footnotesize{}Calculate relative position $r_{AB}$ and relative
orientation $\theta_{AB}$; and get current regime.}{\footnotesize\par}
\item \texttt{\footnotesize{}If in non-interacting regime, apply step 2
from Algorithm \ref{alg:MSM/RDmain}.}{\footnotesize\par}
\item \texttt{\footnotesize{}Else if in transition regime:}{\footnotesize\par}
\begin{enumerate}
\item \texttt{\footnotesize{}Propagate MSM by $\delta t$ using the piecewise
constant approximation of the transition rate matrices $\mathbb{Q}_{AB}$
and $\mathbb{Q}_{AB\rightarrow C}$.}{\footnotesize\par}
\item \texttt{\footnotesize{}If transition towards a bound state $C$ happened
and transition is allowed (bound site not yet occupied):}{\footnotesize\par}
\begin{enumerate}
\item \texttt{\footnotesize{}If both particles don't belong to any compound: create
bound compound with particles $A$ and $B$.}{\footnotesize\par}
\item \texttt{\footnotesize{}Else if only one particle belongs to a compound: bind
particle to existing compound.}{\footnotesize\par}
\item \texttt{\footnotesize{}Else: bind compounds together (particles belong
each to a different compound).}{\footnotesize\par}
\item \texttt{\footnotesize{}Switch to bound regime}{\footnotesize\par}
\item \texttt{\footnotesize{}If new compound was formed: assign it a position
and orientation (can be an average between the particles involved,
or the corresponding center of a ring molecule)}{\footnotesize\par}
\item \texttt{\footnotesize{}For new particles in compound: fix relative
position and orientation with respect to the compound's position and
orientation.}{\footnotesize\par}
\end{enumerate}
\item \texttt{\footnotesize{}Else:}{\footnotesize\par}
\begin{enumerate}
\item \texttt{\footnotesize{}For particles not belonging to a compound: propagate
diffusion by $\delta t$ following Eq. \ref{eq:overdGeneral}}{\footnotesize\par}
\item {\footnotesize{}If }\texttt{\footnotesize{}$\left\Vert r_{AB}\right\Vert \geq R$: switch
to non-interacting regime}{\footnotesize\par}
\end{enumerate}
\end{enumerate}
\end{enumerate}
\texttt{\footnotesize{}For all particle compounds: (bound regime)}{\footnotesize\par}
\begin{enumerate}
\item \texttt{\footnotesize{}Calculate transitions within $\delta t$ in
the compound using $\mathbb{Q}_{C}$ and $\mathbb{Q}_{C\rightarrow AB}$
from the two particle MSM (careful to only calculate one possible
transition per binding).}{\footnotesize\par}
\item \texttt{\footnotesize{}If transition towards an unbound transition
state happened:}{\footnotesize\par}
\begin{enumerate}
\item \texttt{\footnotesize{}Split compound and switch the corresponding
pair of particles to the transition regime.}{\footnotesize\par}
\item \texttt{\footnotesize{}Set position and orientation of particles/compounds
corresponding to the unbound transition state.}{\footnotesize\par}
\end{enumerate}
\item \texttt{\footnotesize{}Else:}{\footnotesize\par}
\begin{enumerate}
\item \texttt{\footnotesize{}Propagate diffusion by $\delta t$ following
Eq. \ref{eq:overdBound} and using the diffusion coefficients for
the corresponding compound (see Appendix \ref{app:diffusionEstimation}).}{\footnotesize\par}
\item \texttt{\footnotesize{}Update the positions and orientations of particles
in compound.}{\footnotesize\par}
\end{enumerate}
\end{enumerate}
\texttt{\footnotesize{}Set}{\footnotesize{} $t\leftarrow t+\delta t$}{\footnotesize\par}
\end{itemize}
\caption{Main algorithm for multiparticle MSM/RD scheme. It is a modified version
of Algorithm \ref{alg:MSM/RDmain}. Particle compounds refer to a
set of particles that are bound together. The compound has its own
position and orientation. The particles positions and orientations
are assigned relative to those of the particle compound. Also note
each possible pair of particles has a regime associated to it. \label{alg:MSM/RDmultipart}}
\end{algorithm*}

\section{Parametrization of the MSM/RD scheme using MD trajectories \label{app:paramMSM/RD}}

The MSM/RD scheme requires parametrizing two MSMs. One for the non-interacting
regime and one for the transition and bound regimes together. The
MSMs for the non-interacting regime can be obtained with standard
techniques \citep{PrinzEtAl_JCP10_MSM1}. In the examples in this
work, the MSMs for the non-interacting regime are trivial since we
already know their form from the benchmark MD model. Thus, we focus
on parametrizing the MSM/RD for the transition and bound regime.

All the MSM/RD schemes in this paper are parametrized with the following
setup: we only allow two molecules in the system, each molecule is
assumed to have a diameter of $5\text{nm}$, which is the same order
of magnitude as some real proteins. The simulation box is a cube with
an edge-length of $25\text{nm}$ and periodic boundaries. Each simulation
runs for $6\times10^{6}$ time steps of $1\text{\ensuremath{\times10^{-5}}}\mu s$
each, yielding a total simulation time of $60\mu s$. We run $600$
of these simulations independently, and we use them to parametrize
the MSM/RD schemes following the steps below.

To parametrize the MSM/RD scheme, we require a piecewise constant
discretization of the transition rate matrix $\mathbb{Q}(x)$. In
the numerical algorithm, it was more convenient to use discrete-time
MSMs; we will need two discrete-time MSMs to approximate $\mathbb{Q}(x)$.
The first one is to approximate the independent dynamics $\mathbb{Q}_{A}\oplus\mathbb{Q}_{B}$
in the non-interacting regime, and the second is to approximate the
dynamics $\mathbb{Q}(x)$ in the bound and transition regime. The
first MSM can be inferred by approximating $\mathbb{Q}_{A}$ and $\mathbb{Q}_{B}$
from MD simulations of molecules $A$ and $B$, each simulated independently
and following standard procedures \citep{PrinzKellerNoe_PCCP11_Perspective,SchererEtAl_JCTC15_EMMA2,SchuetteEtAl_JCP11_Milestoning}.
The second MSM requires discretizing the MD trajectories of the joint
system and then using these discrete trajectories to derive an MSM.
The discretization and derivation is as follows:

\subsubsection*{Discretization of MD trajectories}
\begin{enumerate}
\item If in the non-interacting regime, assign a unique unbound state (we
use zero).
\item If in the transition regime, assign the corresponding transition state,
as defined in Fig.\ref{fig:discretization}b.
\item If in the bound regime, use the core-MSM approach \citep{dibak2018msm,SchuetteEtAl_JCP11_Milestoning}:
if the particles are in a bound state, assign the corresponding bound
state; if they are not in a bound state, assign the value of the last
transition state or bound state visited.
\end{enumerate}

\subsubsection*{Derivation of the MSM}
\begin{enumerate}
\item Slice discrete trajectories by removing the unbound state.
\item Stitch the trajectories randomly by joining a trajectory that ends
in a given state and stitching to another one that begins with that
state.
\item Use the stitched discrete trajectories to generate an MSM that can
transition between all the bound states and all the transition states.
We employ standard inference techniques \citep{PrinzEtAl_JCP10_MSM1},
and we use the PyEMMA software \citep{SchererEtAl_JCTC15_EMMA2} to
obtain the MSM and the corresponding implied time-scales. Observing
the implied time-scales, we can choose an adequate lagtime for the
MSM \citep{PrinzEtAl_JCP10_MSM1}.
\end{enumerate}
The resulting MSM, approximates the full $\mathbb{Q}(x)$ matrix in
the transition and bound regime. The two MSMs derived here can be
used directly with the Algorithm \ref{alg:MSM/RDmain}. In Appendix
\ref{app:paramMSM/RD}, we show the steps to parametrize the MSM/RD
scheme for specific simulations. In addition, the translational and
rotational diffusion coefficients can be obtained from the MD simulations
following standard procedures \citep{kleinhans2005iterative,kutoyants2013statistical,linke2018fully,qian1991single,rao1999statistical,sorensen2004parametric}. 

Below, we show the specific steps to parametrize the MSM/RD scheme
from MD trajectories containing the position and orientation of the
molecules at every $25^{\text{th}}$ time step. We illustrate these
steps for the protein-protein system from Section \ref{subsec:protein-proteinMSMRD},
and we show the minor variations needed for the dimer system from
Section \ref{subsec:implementation1} and for the multiparticle implementation.

\subsection{Protein-protein system:}

To parametrize the rates in the transition regime, we collapsed the
two conformations of $B$ into one state to yield averaged rates over
the two conformations. However, for more complex protein-protein systems,
the framework can handle conformation switching in the transition
regime.
\begin{enumerate}
\item Obtain MD trajectories (position and orientation) from the $600$
simulations with a stride of 25 time steps.
\item Define the six bound states around the position and orientation defined
in Fig. \ref{fig:patchyProteinResults}a.
\item Define bound, transition and unbound regime with $\sigma<\left\Vert r_{AB}\right\Vert <R$;
a reasonable choice for our interaction potential is $\sigma=6.25\text{nm}$
and $R=11.25\text{nm}.$
\item Define transition states in the transition regime by discretizing
the relative position $r_{AB}$ and orientation $\theta_{AB}$ between
the two molecules in the transition regime. Using Fig. \ref{fig:discretization}
as reference, we discretize $r_{AB}$ in $6$ sections and $\theta_{AB}$
into $19$ sections, yielding total of $114$ transition states.
\item Discretize the MD trajectories using the bound states, the transition
states and the unbound state, which is defined if the particles are
in the non-interacting regime. If the particles are in the bound regime
but not in a bound state, use the core MSM approach \citep{SchuetteEtAl_JCP11_Milestoning},
where the state remains the same as the previous state until a new
state is reached.
\item Slice discrete trajectories by removing the unbound state.
\item Stitch the trajectories randomly by joining a trajectory that ends
in a given state and stitching to another one that begins with that
state.
\item Use the stitched discrete trajectories to generate an MSM that can
transition between all the bound states and all the transition states.
We employ standard inference techniques \citep{PrinzEtAl_JCP10_MSM1},
and we use the PyEMMA software \citep{SchererEtAl_JCTC15_EMMA2} to
obtain the MSM and the corresponding implied time-scales. Observing
the implied time-scales, we can choose an adequate lagtime for the
MSM, which in this case is of $150$ data points/steps or $3750$
time steps or $0.0375\mu s,$ see Fig. \ref{fig:ITSboth}a.
\item Incorporate the obtained MSM into the MSM/RD simulation scheme illustrated
in Algorithm \ref{alg:MSM/RDmain}.
\end{enumerate}

\subsection{Dimer system:}

We follow the same steps as in the protein-protein case, except for
steps 4 and 8.

Step 4: Define transition states in the transition regime by discretizing
the relative position $r_{AB}$ and orientation $\theta_{AB}$ between
the two molecules in the transition regime. Using Fig. \ref{fig:discretization}
as reference, we discretize $r_{AB}$ in $7$ sections and $\theta_{AB}$
into $29$ sections, yielding total of $203$ transition states.

Step 8: Same as step 8 before. However the implied time-scales figure
is Fig. \ref{fig:ITSboth}b.

\subsection{Multiparticle system:}

We follow the same steps as in the dimer system, except for step 8.

Step 8: Same as step 8 before. However the implied time-scales figure
is Fig. \ref{fig:ITSboth}c.\medskip{}

\section{MSM/RD algorithms \label{app:MSM/RD-schemes}}

In algorithm \ref{alg:MSM/RDmain}, we present a general outline of
the scheme to simulate MSM/RD. We assume we have already derived a
piecewise constant discretization of the transition rate matrix $\mathbb{Q}(x)$
or of its discrete-time analog in the three regimes. As this is essentially
a numerical scheme to solve Eq. \ref{eq:diffMS_MSMRD}, it is possible
to discretize it in several different ways, yielding each a different
scheme. Nonetheless, they all follow a similar logic. Thus, we present
below a simple scheme; while not necessarily the most accurate nor
the most efficient, it clearly illustrates the logic behind the method.
Note we don't show how to numerically propagate neither the diffusion
nor the MSM. This can be done with several standard methods. In Algorithm
\ref{alg:MSM/RDmultipart}, we further extend the algorithm to handle
multiple particles, as implemented in this paper.

\section{Estimation of diffusion coefficients \label{app:diffusionEstimation}}

As both the benchmark and the MSM/RD simulation use the same model
for diffusion, the diffusion coefficients of compounds are only relevant
for the multiparticle example. To accurately model these multiparticle
compounds with MSM/RD, we estimate the translational and rotational
diffusion coefficients, $D$ and $D^{\mathrm{rot}}$ of the compound
formed by two, three, four and five molecules. We assume anisotropic
rotational diffusion, and we estimate these coefficients by measuring
the mean square displacement at different lag times and applying a
linear fit \citep{qian1991single}. Assuming the position of the center
of the compound at time $t$ is $r(t)$, and its orientation is given
by the quaternion $\theta(t)=\{s,p\}$, on sufficiently long time
scales, the diffusion coefficients follow these relations \citep{qian1991single,linke2018fully}

\begin{align*}
\left\langle \left(r(t+\tau)-r(t)\right)^{2}\right\rangle  & =6D\tau\\
\left\langle \left(p(t+\tau)-p(t)\right)^{2}\right\rangle  & =\frac{3}{4}\left(1-e^{-2D^{\mathrm{rot}}\tau}\right)\approx\frac{3}{2}D^{\mathrm{rot}}\tau.
\end{align*}

There are more accurate and sophisticated methods to estimate the
diffusion coefficients \citep{bullerjahn2020optimal}. For more complex
cases, it is also possible to include anisotropic translational and
rotational diffusion by estimating diffusion matrices \citep{linke2018fully,schluttig2008dynamics}.
However, for our purpose a simple linear fit on these expressions
provides already an accurate result.

\medskip{}

\section*{SI references:}

\begin{btSect}{msmrd-refs,own}
\btPrintCited
\end{btSect}
\end{btUnit}

\end{document}